\documentclass[a4paper]{cas-sc}

\usepackage[authoryear]{natbib}
\usepackage{float}
\usepackage{dirtytalk}
\usepackage{soul}
\usepackage{bm}
\usepackage{graphicx}
\usepackage{amsmath}
\usepackage{tabularx,caption}
\usepackage{multirow}
\usepackage{siunitx}
\usepackage[export]{adjustbox}
\usepackage{microtype}
\usepackage{booktabs}

\def\tsc#1{\csdef{#1}{\textsc{\lowercase{#1}}\xspace}}
\tsc{WGM}
\tsc{QE}
\tsc{EP}
\tsc{PMS}
\tsc{BEC}
\tsc{DE}

\begin{document}
\let\WriteBookmarks\relax
\def\floatpagepagefraction{1}
\def\textpagefraction{.001}

\title [mode = title]{Bond exchange reactions as a paradigm for mitigating residual stress in polymer matrix fiber composites}   
   
\shorttitle{BERs to mitigate res. stress}

\author[1]{Zhongtong Wang}
\credit{Conceptualization, methodology, software, formal analysis, visualization, writing – original draft}

\author[1]{ Robert J. Wagner}
\credit{Conceptualization, methodology, software, formal analysis, writing-review and editing}

\author[2]{ Tianke Chen}
\credit{Methodology, writing-review and editing}

\author[3]{ Sagar P. Shah}
\credit{Writing-review and editing}

\author[3]{ Marianna Maiaru}
\credit{Conceptualization, Writing-review and editing}

\author%
[1]{ Meredith N. Silberstein}
\cormark[1]
\credit{Conceptualization, methodology, project administration, supervision, writing-review and editing, and funding acquisition}
\affiliation[1]{organization={Sibley School of Mechanical and Aerospace Engineering, Cornell University},
    city={Ithaca},
    postcode={14853, NY}, 
    country={U.S.A.}}

\affiliation[2]{organization={Department of Materials Science and Engineering, Cornell University},
    city={Ithaca},
    postcode={14853, NY}, 
    country={U.S.A.}}

 \affiliation[3]{organization={Department of Civil Engineering and Engineering Mechanics, Columbia University},
    city={New York},
    postcode={10027, NY}, 
    country={U.S.A.}}    

\cortext[cor1]{Correspondence to Meredith N. Silberstein. Email: meredith.silberstein@cornell.edu}

\begin{abstract}
Polymer matrix fiber composites often suffer from residual stresses due to differences in coefficients of thermal expansion between the fibers and resins, as well as contractile strain of the resins during curing. To address residual stress driven composite failure, we propose the use of vitrimers as composite resins, which can undergo thermally activated, stress alleviating, bond exchange reactions (BERs).  We conduct fiber Bragg grating measurements for a single glass fiber within bulk vitrimer. These show that the fiber strain in vitrimers with 5$\%$ catalyst is significantly lower than in those with  0$\%$ catalyst (minimal BER expected) during both curing and post-curing phases. We developed a finite deformation, micromechanically-inspired model that incorporates curing, thermal processes, and BERs, and then implemented this model it into finite element software to simulate stress evolution within single fiber composite systems. The combination of experimental and computational results reveals that BERs can effectively mitigate, but not eliminate, the residual stress in polymer matrix fiber composites. 

\end{abstract}


\begin{highlights}
\item Propose that using vitrimers with bond exchange reactions can relieve residual stresses within fiber composites.
\item Devised a rheological model that encompasses curing, thermal processes and bond exchange reactions.
\item Both fiber Bragg grating experiments and FEM simulations indicate that bond exchange reactions can reduce the residual stress in polymer matrix fiber composites. 
\end{highlights}


\maketitle

\section{Introduction}

Polymer matrix fiber composites are engineered materials in which high-strength fibers such as carbon and glass are embedded within a polymer matrix to create a composite that leverages the mechanical benefits of both components. The fibers provide reinforcement, contributing tensile strength and rigidity, while the matrix distributes loads among fibers and offers shape and environmental resistance \citep{herrera2005study,hyer2009stress,kang2014modeling,sharma2023critical,zheng2022recent}. Polymer matrix fiber composites are prevalent across various fields such as automotive, electronics, consumer goods, construction and infrastructure \citep{prashanth2017fiber,qureshi2022review,duflou2012fiber}. For example, polymer matrix fiber composites are utilized for structural components and body panels to reduce weight and improve fuel efficiency in the automotive industry. In electronics, polymer matrix fiber composites are employed in the fabrication of high voltage switches, cryostats, and dry transformers.

The composite manufacturing process is a critical step in the design and development of polymer matrix fiber composites structural components as it governs both the final geometry of the manufactured polymer matrix fiber composites part as well as its mechanical performance \citep{baran2017review}. The manufacturing process includes curing, an exothermic chemical reaction, during which the matrix material properties evolve as a function of time and temperature. During manufacturing, the thermal expansion coefficient mismatch between the constituent fibers and matrix leads to differential temperature-induced volumetric strains within the composite microstructure. These strains, in combination with the chemical volumetric shrinkage manifested by the matrix material during cure and its thermo-mechanical property evolution, result in self-equilibrating, performance-altering residual stress generation \citep{baran2017review,mesogitis2014uncertainty,shah2023process}. At the constituent scale, these process-induced residual stresses can lead to the formation of defects such as micro-cracks, delamination, and voids within the matrix, significantly reducing the composite mechanical performance and service life \citep{hu2018investigation,sorrentino2017new,d2016virtual}. At the structural scale, residual stresses can induce structural warping, resulting in dimensional changes that impact geometrical tolerances and assembly \citep{fernlund2002experimental, chen2019improved, nagaraj2024validation}. Therefore, it is crucial to address the process-induced residual stress generation in polymer matrix fiber composites during manufacturing, in order to fabricate composite parts with more confidence and exploit their full potential. 

Over the past few decades, researchers have developed various strategies focused on physics-based multiscale process modeling to highlight the underlying mechanisms that lead to residual stress generation and quantify their influence on the composite part shape and performance \citep{fish2021mesoscopic,llorca2011multiscale, danzi2019numerical, d2017virtual,he2019multiscale, hui2021integrated, liu2013computational, shah2020multiscale}. Such computational process models predict the residual stresses and provide opportunities for cure cycle optimization to alleviate their influence on the composite response. Furthermore, several researchers have reported strategies that modify the processing techniques and composite design to account for residual stresses. For example,  applying tensile stress to reinforcement fibers during curing can neutralize the contraction of the resin, effectively reducing residual stress \citep{mostafa2017fibre,mohamed2020development, schlichting2010composite,fazal2014uhmwpe}. Preheating the composite mold to the curing temperature leads to more uniform samples with less residual stress, due to the even distribution of heat \citep{zhang2014effects,zhang2014effect,tabatabaeian2022residual}. Another approach involves careful temperature control during curing to minimize residual stress, such as the modified cure cycle comprising abrupt cooling after gelation reduces the residual stresses because an abrupt-cooling operation after gelation could efficiently dissipate the strain generated by the laminate thermal contraction due to the resin’s viscoelastic behavior \citep{kim2012study,chava2022residual}. Innovative composite designs and laminate scheduling, particularly for complex shapes like T-joints \citep{burns2016strengthening}, help distribute stresses more evenly. Furthermore, the incorporation of nano-additives has emerged as a promising method for mitigating residual stress. Research indicates that adding carbon nanofibers to an epoxy matrix significantly reduces its coefficient of thermal expansion (CTE) to more closely match that of the primary structural reinforcement fibers, while moderately enhancing its Young's modulus \citep{shokrieh2014reduction,ghasemi2015role,tabatabaeian2022residual}.  While the aformentioned methods are effective in reducing residual stress in polymer matrix composites, they introduce complexity to manufacturing processes. Furthermore, due to the permanent chemical crosslinking structures of the polymer matrices used, which are typically thermosets, additional processing steps to alleviate residual stress become unavailable after cure completion. Therefore, residual stress states become locked in and unadjustably linked to temperature after initial cure so that manufacturers cannot correct thermal processing errors or modify operational temperatures. These challenges have spurred ongoing research aimed at exploring modifications to the matrix material itself as a strategy for alleviating residual stress. 

A particularly promising class of materials for use as post-processable epoxy matrices is vitrimers. Vitrimers are derived from and mechanically similar to thermosets, but are cross-linked by dynamic covalent bonds that can reversibly undergo bond exchange reactions (BERs) \citep{montarnal2011silica,denissen2016vitrimers,rottger2017high,winne2019dynamic} that allow them to reach more thermodynamically stable states. These BERs can be triggered by various external stimuli, such as exposure to UV light \citep{ma2014photoviscoplastic,liu2021biobased} or an increase in temperature \citep{shi2021nonequilibrium,liu2018self,hubbard2021vitrimer}. Thermally activated vitrimers exhibit the robust thermal and mechanical properties of traditional thermosets when well below their topology freezing (or vitrification) transition temperature ($T_v$). Above $T_v$ however, they undergo rapid exchangeable reactions, such as transesterification, facilitating the material's ability to reconfigure. Therefore, by replacing thermoset matrices with vitrimer counterparts, new capabilities emerge such as healability \citep{zhao2019self,yang2018solvent}, weldability \citep{yu2016interfacial,an2022chain}, and recyclability \citep{taynton2016repairable,shi2017recyclable}. For example, after immersing a carbon fiber reinforced vitrimeric polymer composites in solvent and increasing the temperature, researchers were able to dissolve the vitrimeric matrix and reclaim clean carbon fibers \citep{yu2016carbon}.

In the present work, we propose using vitrimers with BERs as an innovative solution to relieve residual stresses within fiber-reinforced composites. First, we introduce a micromechanically inspired, three-dimensional model to capture the main mechanical behaviors relevant to curing and thermal cycling of vitrimers. These are temperature-dependent viscoelasticity; temperature-induced and chemical shrinkage-induced volumetric changes; and temperature-dependent BER-derived relaxation (Section \ref{sec:model}). Subsequently, we discuss experimental characterization procedures and finite element approach used to capture the neat vitrimer behavior, as well as the procedure to measure and simulate a composite consisting of a single fiber within a vitrimer (Section \ref{sec:methods}). We examine two formulations of the vitrimer: 5$\%$ catalyst as the vitrimer with significant BERs and 0$\%$ catalyst as the control with no BER. We then present experimental and simulation results for the neat vitrimer and single fiber composite (Section \ref{sec:results}). We look separately at high temperature curing, cooling to room temperature, and post-cure thermal cycling to understand the potential for BER utility in relieving residual stress.

\section{Model framework} \label{sec:model}

   \begin{figure}[pos=h]
	\centering
		\includegraphics[scale=.28]{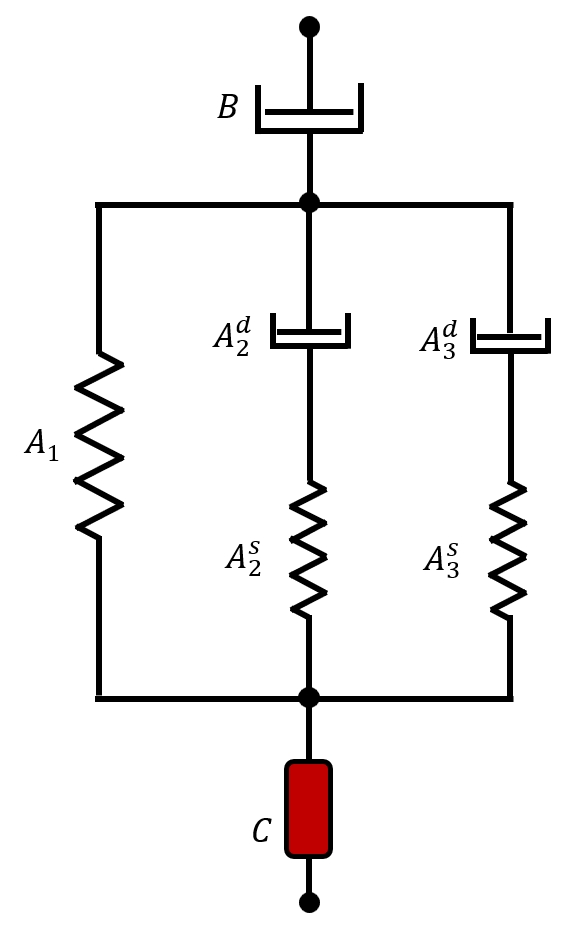}
	      \caption{ A 1D representation of the proposed 3D finite deformation rheological model.}
	\label{fig:schematic}
    \end{figure}

Here we formulate a three dimensional finite deformation model for vitrimers that utilize elevated temperature to activate BERs.  For thermally activated vitrimers, it is crucial to differentiate between two types of viscous deformation: one resulting from the thermally driven changes in chain conformations and the other from BERs. Given the focus of our study on residual stress arising during composites processing, the model we formulate captures temperature dependent viscoelasticity, volumetric changes resulting from changes in temperature and curing, and temperature dependent BERs-derived relaxation. 

A one-dimensional rheological representation of the constitutive model is shown in Figure \ref{fig:schematic}. This model follows many of the conventions in the field of glassy polymer continuum modeling, and specifically builds on the work of \citet{ma2014photoviscoplastic} for polymers with light triggered BERs and \citet{mao2019viscoelastic} shape memory of polymers with thermally triggered BERs. This model is comprised of seven key rheological elements. Element $C$ captures both cure strain and thermal strain (Section \ref{sec:element1}). Viscoelastic element $A$ includes the elements $A_1$, $A^s_2$, $A^d_2$, $A^s_3$ and $A^d_3$ to capture temperature dependent linear viscoelasticity (Section \ref{sec:element2}). To capture the two distinct relaxation timescales of the vitrimer, we employ two Maxwell elements ($A_2$, $A_3$) in our modeling approach. Element $B$ captures relaxation due to thermally activated BERs (Section \ref{sec:element3}).

The total deformation gradient can be decomposed into three contributions as: 

\begin{align}
    &\textbf{F} = \textbf{F}_A \textbf{F}_B \textbf{F}_{C}
\end{align}
\noindent where $\textbf{F}_A$ is the deformation gradient of the total viscoelastic element,  $ \textbf{F}_B$ is the relaxation deformation due to BERs-enabled flow, and $\textbf{F}_{C}$ is the deformation gradient imposed by curing and temperature changes. The deformation gradients within the viscoelastic branch can be expressed as: 
\begin{align}
    &\textbf{F}_A = \textbf{F}_{A1} =  \textbf{F}_{A_2^s}\textbf{F}_{A_2^d} =  \textbf{F}_{A_3^s}\textbf{F}_{A_3^d}
\end{align}
\noindent where $\textbf{F}_{A1}$ is the deformation gradient of equilibrium spring $A_1$; and $\textbf{F}_{A_2^s}$, $\textbf{F}_{A_2^d}$, $\textbf{F}_{A_3^s}$ and $\textbf{F}_{A_3^d}$ are the deformation gradients of the elements $A^s_2$, $A^d_2$, $A^s_3$ and $A^d_3$ respectively.

The Cauchy stress ($\textbf{T}$) of the system is equal to the stress on the viscoelastic component ($\textbf{T}_{A}$), as well as the stress on the BER element ($\textbf{T}_{B}$).

\begin{align}
    & \textbf{T} = \textbf{T}_{A} = \textbf{T}_{B}
\end{align}

The stress on the element $A$ is given by the sum of the legs, as:

\begin{align}
    & \textbf{T}_{A} = \textbf{T}_{A1}+ \textbf{T}_{A2} +\textbf{T}_{A3}
\end{align}

\noindent where $ \textbf{T}_{A1}$,$ \textbf{T}_{A2}$ and $\textbf{T}_{A3}$ are the contributions to the true stress originating from the equilibrium spring $A_1$, linear spring $A^s_2$ and linear spring $A^s_3$ respectively.

\subsection{Element C - volumetric expansion/contraction element} \label{sec:element1}

As highlighted in the introduction, residual stress occurs during the curing process from both the curing itself and from the thermal strain during subsequent cooling. These strains significantly impact the post-cure properties of the composite. During the curing process of the vitrimer matrix, we adopt the simple assumption \citep{adolf1998stresses,shah2023process} that the cure strain, $\epsilon_c$, is proportional to the degree of cure, $\phi$, as: 
\begin{align}
   &\epsilon_c=  -\alpha_c \phi
\end{align}   
\noindent where $\alpha_c $ is a cure contraction coefficient. The rate of cure is assumed to scale directly with the proportion of unreacted material as:
\begin{align}
    &\frac{d\phi}{dt}  =  \frac{1}{\tau_{\phi}} \left(1- \phi \right)
\end{align}
\noindent where $\tau_\phi^{-1}$ is a characteristic cure rate and initial cure is taken as $\phi(t=0)=0$. The cure strain coefficient, $\alpha_c$, and the cure rate coefficient, $\tau_\phi^{-1}$, are material specific parameters. In general, $\tau_{\phi}$ is temperature dependent, but here we will use a single cure temperature, rendering this a constant.

Thermal strain ($\epsilon_{T}$) in the vitrimer is estimated using a linear piecewise equation, which captures its evolution across distinct thermal regimes: glassy, elastomeric, and vitrimeric. Each regime is characterized by a unique CTE, denoted as $\alpha_{nT}$.  

The deformation gradient for the element $C$ is assumed isotropic and calculated using the cure strain and thermal strain as:

\begin{align}
    &\textbf{F}_C = \exp\left(\epsilon_c\right) \exp\left(\epsilon_T\right) \textbf{I} 
\end{align}

\noindent where $\textbf{I}$ is the 3x3 identity tensor.

\subsection{Element A - viscoelastic element} \label{sec:element2}

The viscoelastic element models the temperature-dependent viscoelastic behavior of the material. It comprises a linear equilibrium spring, denoted as $A_1$, arranged in parallel with two Maxwell element. The first Maxwell element, $A_2$, consists of a linear elastic spring ($A^s_2$) in series with a dashpot ($A^d_2$). Similarly, the second Maxwell unit, $A_3$, includes a linear elastic spring ($A^s_3$) in series with a dashpot ($A^d_3$). The true stress of the linear spring element $A_1$ can be calculated as: 
\begin{align}
    &\textbf{T}_{A1} = \mathcal{L}_{A1}( \ln \textbf{V}_{A1})
\end{align}

\noindent where $\textbf{V}_{A1}$ is the left stretch tensor of the linear spring $A_1$ obtained from the polar decomposition. $\mathcal{L}_{A1}$ is the fourth order isotropic elasticity tensors defined as:

\begin{align}
    &\mathcal{L}_{A1} = 2 \mu_{A1} \mathcal{I} + \left(\kappa_{A1} - \frac{2}{3} \mu_{A1}\right) \textbf{I} \otimes \textbf{I} 
\end{align}

\noindent where $\mu_{A1}$ and $\kappa_{A1}$ are the shear and bulk modulus for linear spring $A_1$, respectively.

The true stress of the first Maxwell unit $A_2$ can be calculated as: 
\begin{align}
    &\textbf{T}_{A2} = \mathcal{L}_{A2}(\ln \textbf{V}_{A^s_2})
\end{align}

\noindent where $\textbf{V}_{A^s_2}$ is the left stretch tensor of the linear spring $A^s_2$ obtained from the polar decomposition. $\mathcal{L}_{A2}$ is the fourth order isotropic elasticity tensor defined as:

\begin{align}
    &\mathcal{L}_{A2} = 2 \mu_{A2} \mathcal{I} + \left(\kappa_{A2} - \frac{2}{3} \mu_{A2}\right) \textbf{I} \otimes \textbf{I} 
\end{align}

\noindent where $\mu_{A2}$ and $\kappa_{A2}$ are the shear and bulk modulus for the Maxwell unit $A_2$ leg respectively.

Similarly, the true stress of the second nonequilibrium branch $A_3$ can be expressed as: 
\begin{align}
    &\textbf{T}_{A3} = \mathcal{L}_{A3}(\ln \textbf{V}_{A^s_3})
\end{align}

\noindent where $\textbf{V}_{A^s_3}$ is the left stretch tensor of the linear spring $A^s_3$ obtained from the polar decomposition. $\mathcal{L}_{A3}$ is the fourth order isotropic elasticity tensor calculated as:

\begin{align}
    &\mathcal{L}_{A3} = 2 \mu_{A3} \mathcal{I} + \left(\kappa_{A3} - \frac{2}{3} \mu_{A3}\right) \textbf{I} \otimes \textbf{I} 
\end{align}

\noindent where $\mu_{A3}$ and $\kappa_{A3}$ are the shear and bulk modulus of the linear spring $A^s_3$ respectively.

In order to capture the temperature dependent response of the system, we treat the Young's modulus as a fitted functional of temperature.    Since the temperature dependent Young's modulus is typically easier to measure experimentally for these materials than the shear modulus, we can also calculate these elastic constants as: 
\begin{align}
    & \mu_{A1}\left(T\right) = \frac{E_{A1}\left(T\right)}{2(1+\nu)} 
    & \kappa_{A1}\left(T\right) = \frac{1}{3(1-2\nu)} 2\mu_{A1}\left(T\right) (1+\nu)  \\
    &\mu_{A2}\left(T\right) = \frac{E_{A2}\left(T\right)}{2(1+\nu)}
    &\kappa_{A2}\left(T\right) = \frac{1}{3(1-2\nu)} 2\mu_{A2}\left(T\right) (1+\nu)  \\
    &\mu_{A3}\left(T\right) = \frac{E_{A3}\left(T\right)}{2(1+\nu)}
    &\kappa_{A3}\left(T\right) = \frac{1}{3(1-2\nu)} 2\mu_{A3}\left(T\right) (1+\nu)     
\end{align}

\noindent where $\nu$ is the Poisson's ratio, assumed to be constant, and $E_{A1}\left(T\right)$, $E_{A2}\left(T\right)$ and $E_{A3}\left(T\right)$ are the elastic modulus for element $A_1$, $A^s_2$ and $A^s_3$ respectively. The detailed compression relaxation experiment and identification process for temperature dependent modulus are shown in Appendix \ref{app:compress}.

The evolution of flow for the linear dashpot $A^d_2$ is given by
\begin{align}
    &\dot{\textbf{F}}_{A^d_2} = \textbf{D}_{A^d_2}\textbf{F}_{A^d_2} 
\end{align}

\noindent where ${\textbf{F}}_{A^d_2}$ is the deformation gradient for element $A^d_2$, $\dot\square$ represents the material time derivative, and ${\textbf{D}}_{A^d_2}$ is the rate of viscous stretching. The flow rule of the viscoelastic component is prescribed as: 

\begin{align}
    &\textbf{D}_{A^d_2} = \frac{\dot{\gamma}_{A^d_2} \textbf{T}^s_{A2}}{2\sigma_{A2}} \label{eq:viscous_stretch} \\
    &\dot{\gamma}_{A^d_2} = \frac{\sigma_{A2}}{\mu_{A2} \left(T\right) \tau^T_{2}\left(T\right) } \\
    &\sigma_{A2} = \sqrt{\frac{\textbf{T}^s_{A2}:\textbf{T}^s_{A2}}{2}}
\end{align}

\noindent where $\textbf{T}^s_{A2} =\textbf{T}_{A2} - \frac{1}{3} tr\left(\textbf{T}_{A2}\right)\textbf{I} $  is the deviatoric stress acting on the viscoelastic component, $\dot{\gamma}_{A^d_2}$  is the strain rate, and $\sigma_{A2}$ is the equivalent shear stress for the first nonequilibrium branch $A_2$. Thermal relaxation timescale $\tau^T_2\left(T\right) $ is a fit function of temperature identified in Appendix \ref{app:compress}.

The evolution of flow for the linear dashpot $A^d_3$ is expressed by:
\begin{align}
    &\dot{\textbf{F}}_{A^d_3} = \textbf{D}_{A^d_3}\textbf{F}_{A^d_3} 
\end{align}

\noindent where ${\textbf{F}}_{A^d_3}$ is the deformation gradient and ${\textbf{D}}_{A^d_3}$ is the rate of viscous stretching for dashpot $A^d_3$. The flow rule of the dashpot $A^d_3$ is prescribed as: 

\begin{align}
    &\textbf{D}_{A^d_3} = \frac{\dot{\gamma}_{A^d_3} \textbf{T}^s_{A3}}{2\sigma_{A3}}  \\
    &\dot{\gamma}_{A^d_3} = \frac{\sigma_{A3}}{\mu_{A3} \left(T\right) \tau^T_{3}\left(T\right) } \\
    &\sigma_{A3} = \sqrt{\frac{\textbf{T}^s_{A3}:\textbf{T}^s_{A3}}{2}}
\end{align}

\noindent where $\textbf{T}^s_{A3} =\textbf{T}_{A3} - \frac{1}{3} tr\left(\textbf{T}_{A3}\right)\textbf{I} $  is the deviatoric stress acting on the viscoelastic component, $\dot{\gamma}_{A^d_3}$  is the strain rate, and $\sigma_{A3}$ is the equivalent shear stress for the first nonequilibrium branch $A_3$. $\tau^T_3\left(T\right) $ is thermal relaxation timescale as a fit function (Appendix \ref{app:compress}).

\subsection{Element B - viscous BER flow} \label{sec:element3}

As mentioned previously, the macroscopic response of vitrimer to elevated temperature manifests as viscoelastic-like behavior, characterized by viscous deformation and associated stress relaxation. To model the viscous deformation due to BERs, a dashpot labeled B is employed in the framework. This dashpot specifically accounts for the viscous deformation linked to the dynamic rearrangement of the network under thermal stimuli. The evolution of the flow for the dashpot B is expressed by:
\begin{align}
     \dot{\textbf{F}}_B = \textbf{D}_{B}\textbf{F}_{B} 
\end{align}
\noindent where ${\textbf{F}}_{B}$ is the deformation gradient and ${\textbf{D}}_{B}$ is the rate of viscous stretching for BER element $B$. The flow rule of the BER element $B$ is given by:

\begin{align}
    &\textbf{D}_{B} = \frac{\dot{\gamma}_B \textbf{T}^s}{2\sigma_B}  \\
    &\dot{\gamma}_B = \frac{\sigma_B}{\mu_A \left(T\right) \tau_{B} \left(T\right)} \\
    & \sigma_B = \sqrt{\frac{\textbf{T}^s:\textbf{T}^s}{2}}
\end{align}

\noindent where $\textbf{T}^s=\textbf{T} - \frac{1}{3} tr\left(\textbf{T}\right)\textbf{I}$ is the deviatoric part of the total Cauchy stress, and $\mu_{A}\left(T\right)$ is the effective shear modulus for this material in the absence of BERs, for simplicity, here we take $\mu_A \left(T\right) = \mu_{A1} \left(T\right) +\mu_{A2} \left(T\right) +\mu_{A3} \left(T\right)$. $\dot{\gamma}_{B}$  is the shear strain rate for element $B$ and $\sigma_{B}$ is the equivalent shear stress for the total system.

The timescale, $\tau_{B}$, is the characteristic time of BER-induced topology transition. We assume that the BER relaxation timescale follows the time-temperature superposition principle,
\begin{align}
    &\tau_B(T) = \tau_B^0 \alpha_B(T) 
\end{align}
\noindent where $\tau_B^0$ is reference relaxation time. Since the BER relaxation time $\tau_{B}$ is related to the kinetics of BERs \citep{shi2021nonequilibrium}, Arrhenius’ law is applied for the shift factor $\alpha_B(T)$
\begin{align}
    & \ln \alpha_B(T) = \frac{E_a}{R}\left(\frac{1}{T}-\frac{1}{T_0}\right)
\end{align}
\noindent where $E_a$ is the BER activation energy which is determined as a fitting parameter, $R$ is gas constant and $T_0$ is the reference temperature for BER flow.

\section{Methods} \label{sec:methods}

\subsection{Vitrimer Synthesis}

We selected a previously studied epoxy-based vitrimer for the experimental work and to drive specification of the model \citep{hubbard2021vitrimer,hubbard2022creep}. The vitrimer was synthesized from Bisphenol A diglycidyl ether (DGEBA) with sebacic acid (SA, 99$\%$) and 1,5,7-triazabicyclo[4.4.0]dec-5-ene (TBD) \citep{hubbard2021vitrimer}. All materials were purchased from Sigma Aldrich and used as received. The synthesis was carried out by using a 1:1 molar ratio of DGEBA and SA with 0 or 5 mol$\%$ TBD as catalyst. The DGEBA was degassed in vacuum for one hour while sitting in an 80$^\circ$C oil bath. Then SA was added, and the temperature was elevated to 140$^\circ$C. The mixture was stirred at 300 rpm and degassed in vacuum for half an hour until the SA was completely melted and homogeneously mixed. During this time, silicone molds (either to produce bulk vitrimer samples for material testing or for single fiber composite samples) were preheated in the vacuum oven at 180$^\circ$C. After mixing in SA, TBD was weighed, poured, and stirred into the mixture (at 300 rpm for one minute). Finally, the resin was poured into the preheated molds and cured in an oven at 180$^\circ$C for 24 hours. From literature we expect that material with 0$\%$ catalyst and 5 mol$\%$ catalyst will have a glass transition temperature of approximately 53$^\circ$C and 36$^\circ$C respectively \citep{hubbard2021vitrimer}.

\subsection{Single-fiber Fiber Bragg Experiments}

\begin{figure}[pos=h]
	\centering
		\includegraphics[scale=.3]{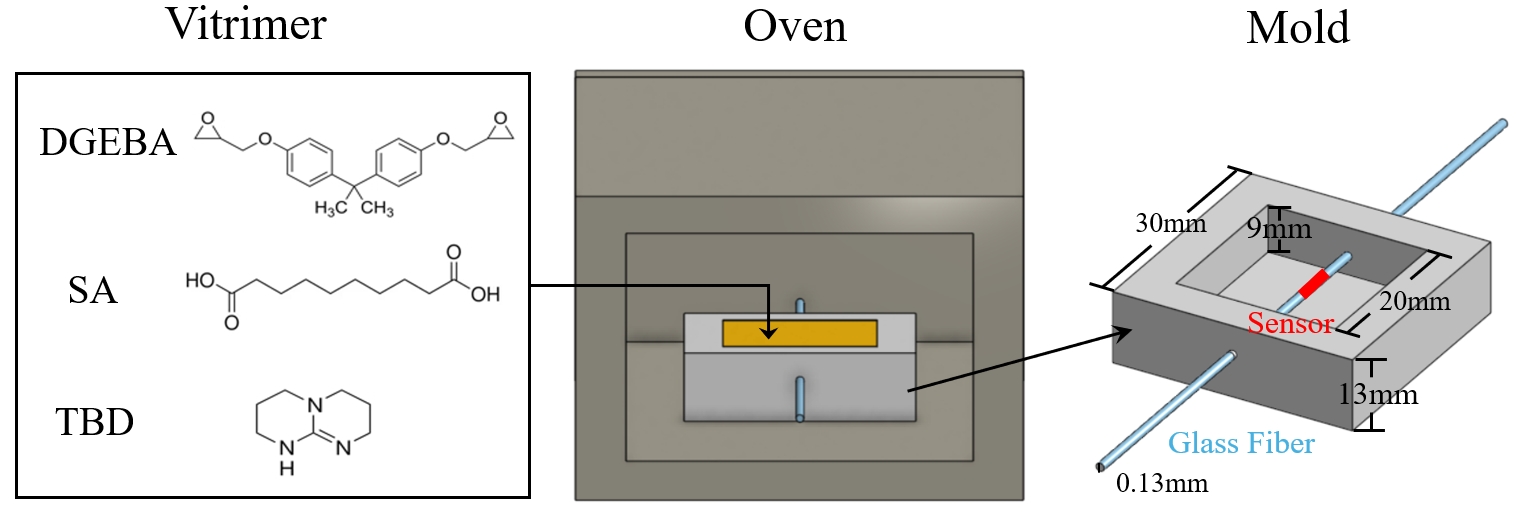}
	      \caption{ The vitrimer was synthesized using DGEBA, SA, and TBD. The sensor of the FBG glass fiber, marked in red, was positioned within a silicone mold. This mold was then placed in an oven, and the vitrimer (yellow) was poured into it to begin the curing process. }
	\label{fig:expsetup}
    \end{figure}

A Fiber Bragg Grating (FBG) sensor system was used to measure residual stress in our vitrimer fiber composite. The diameter of the fiber is 0.13mm. The sensor end of the FBG fiber (T20 from Technica) was placed in a silicone mold with dimension of 9mm $\times$ 20mm $\times$ 20mm (size selected to minimize edge effects) and was preheated in a 180$^\circ$C oven for 1 hour until the vitrimer is mixed. The other end of the FBG fiber was connected to a M4 interrogator system from Technica. The pre-cured vitrimer with 0$\%$ and 5$\%$ catalyst concentration was then poured into the mold on top of the FBG fiber. The vitrimer was then cured at an oven temperature of 180$^\circ$C for 24 hours as shown in Figure \ref{fig:expsetup}. After curing, the sample was cooled at 5$^\circ$C/min to 25$^\circ$C and was held at 25$^\circ$C for 45 minutes to monitor the relaxation at low temperature. After this, the oven temperature was raised to 220$^\circ$C at 5$^\circ$C/min and was held there for 1 hour. Lastly, the oven temperature was again cooled to 25$^\circ$C at 5$^\circ$C/min and held for 45 minutes to monitor the strain difference before and after the BER. The FBG sensor reading was recorded throughout curing and thermal cycling and its spectrum peak value was converted to a strain value using the FBG manufacturer provided conversion constant ($8.33\times 10^{-4} $ nm $^{-1}$). Temperature was measured throughout experiments using a thermocouple immediately adjacent to the specimen.

\subsection{Compression Relaxation Experiments}

Compression relaxation experiments were conducted using a ZwickRoell Z010 universal testing machine equipped with a 10 kN load cell at temperatures of 21$^\circ$C, 35$^\circ$C, 50$^\circ$C, 70$^\circ$C, 100$^\circ$C, 120$^\circ$C, and 170$^\circ$C. Cylindrical specimens with a diameter of 9.56 mm, a height of 13.76 mm, and 0$\%$ catalyst  were prepared for testing. A loading rate of 120 mm/min was applied. To ensure alignment, a pre-load of 1 N was applied to each specimen at the beginning of each test. Once the target temperature was reached, the specimen was held for 4 hours to achieve thermal equilibrium. For experimental temperatures below 50°C, the sample was compressed to a strain of 3$\%$.  Then the deformation was maintained during the following 15 hours. The resulting stress decay was recorded to calculate the relaxation modulus. For temperatures at or above 50°C, a strain of 8$\%$ was applied since the linear elastic region is larger above the glass transition temperature. Due to the rapid relaxation observed at 70°C and above, we focused our analysis solely on the uniaxial compression region for these higher temperatures and assumed arbitrary ``fast" relaxation constants of 10s. 

\subsection{Thermomechanical Analysis (TMA) Experiments} \label{sec:TMA}

TMA was conducted using a TA Instruments Q400EM to measure the thermal expansion of the vitrimer as a function of temperature. Samples were prepared with dimensions of 8 mm$\ \times\ $8 mm$\ \times\ $5 mm. A feedback force of 0.01 N was applied to the compression probing tip resting on top of the vitrimer to ensure consistent contact. The sample was heated from 0$^\circ$C to 220$^\circ$C at a rate of 5$^\circ$C min$^{-1}$, held at 220$^\circ$C to reach thermal equilibrium, and then cooled back down to 0$^\circ$C at the same rate. CTE were extracted from the cooling phase of the experiment.

\subsection{Simulation of the Single Fiber Composite system} \label{sec:sim}


    \begin{figure}[pos=h]
	\centering
		\includegraphics[scale=.2]{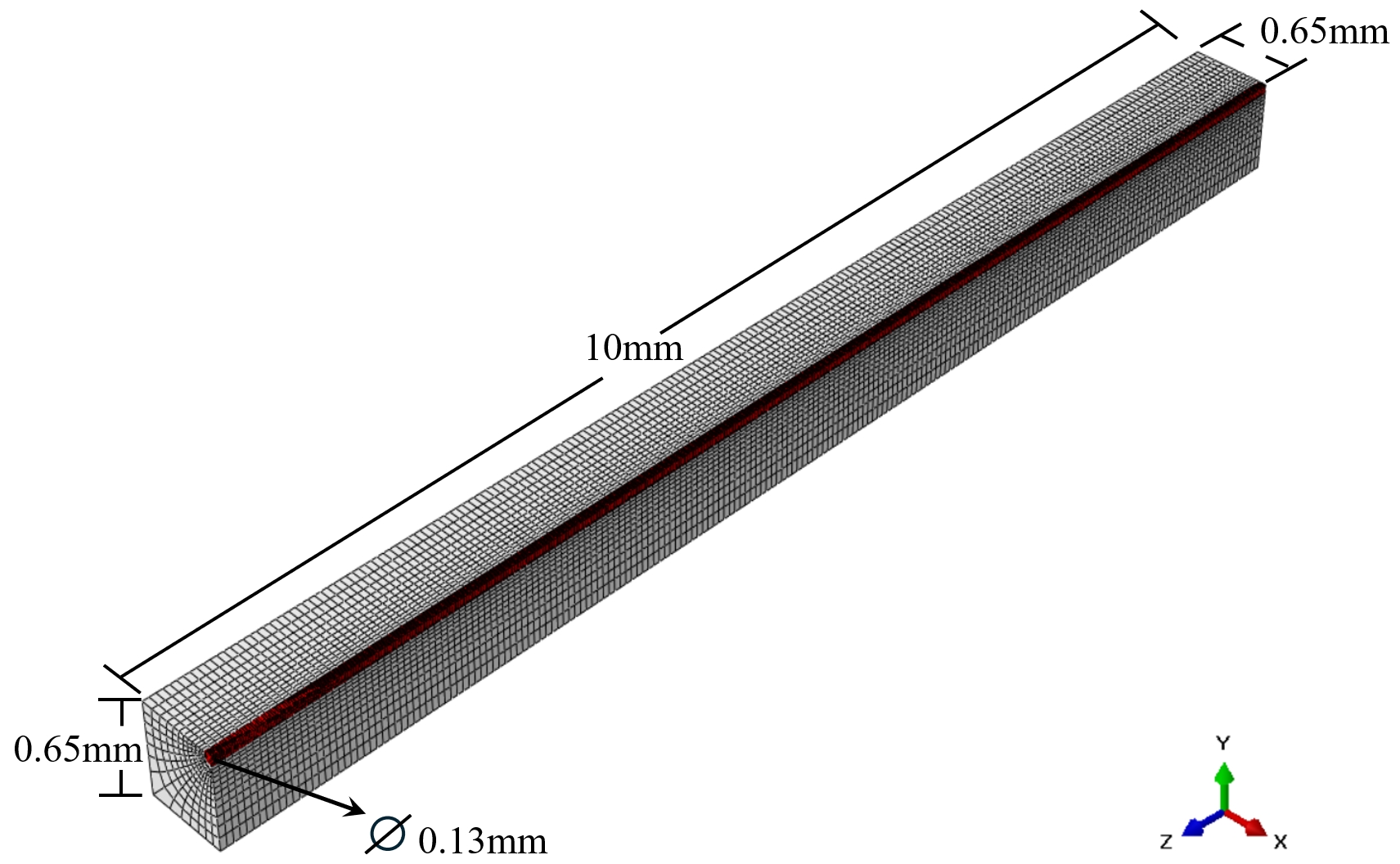}
	      \caption{The FEA mesh is shown with the glass fiber in red and the vitrimer in off-white. The central axis of the glass fiber aligns with the z-axis of the model domain. For computational efficiency a one-eighth-space model is used with planar symmetry about the xy-, yz-, and zx-planes.}
	\label{fig:abaqus}
    \end{figure}
    
A single fiber composite analogous to that of the fiber Bragg experiments was simulated within the finite element software Abaqus STANDARD. This model was used for multiple purposes, which were to extract material properties, to interpret residual stresses within the vitrimer consistent with the experimental fiber strain results, and to make predictions about the influence of thermal cycling on residual stresses. The vitrimer constitutive model was implemented into a user material subroutine (UMAT) that employs the two-step iterative implicit integration scheme put forth by \citet{ma2014photoviscoplastic}. The details of the implicit integration scheme are documented in Appendix \ref{app:scheme}.  The glass fiber was captured using the built-in, isotropic linear thermoelastic material model of Abaqus. The model geometry is depicted in Figure \ref{fig:abaqus}, where the off-white domain is vitrimer matrix and the red domain is the glass fiber. The fiber is circumferentially enveloped by vitrimer with perfect bonding assumed at the interface. A one-eighth model was used to reduce computational cost wherein the longitudinal axis of the fiber aligns with the z-axis of the domain. The geometry is planar symmetric about the xy-, yz-, and zx-planes. The three, non-planar-symmetric faces of the model are traction-free boundaries. The outer dimension for the vitrimer in Abaqus are 0.65 mm $\times$ 0.65 mm $\times$ 10 mm. The diameter for the glass fiber was set to 0.13 mm to match the experimental setup. The distance from the longitudinal axis to the traction-free faces of the vitrimer was set such that the stress profile near the fiber is unaffected by the boundaries (i.e., a far-field approximation is achieved). In our finite element analysis, we employed C3D8 elementsand varied the mesh size to ensure solution convergence. The model parameter values, as well as the associated experimental identification methods, are listed in Table \ref{tab:parameter}. Details of the fitting procedures are given in Appendix \ref{app:para_ident}.

\begin{center}
\captionof{table}{Simulation parameters} \label{tab:parameter}
\begin{tabular}{llll} 
\toprule
{Parameter} &  \multicolumn{2}{c}{Determined values} & {Description}  \\ 
\midrule

Matrix&  0$\%$ Vitrimer& 5$\%$ Vitrimer & {}\\

$\alpha_c$ &  0.0077 & 0.0077   & Cure contraction coefficient \\ 

$\tau_\phi^{-1}$ & 1.75e-5 s$^{-1}$ &  1.75e-5 s$^{-1}$& Cure rate \\

$\alpha_{nT}$ &  1.761e-4 $^\circ $C$^{-1}$ &  1.765e-4 $^\circ $C$^{-1}$ & $T<T_0$  thermal expansion \\ 
 &  2.486e-4 $^\circ $C$^{-1}$&  2.291e-4$^\circ $C$^{-1}$  & $T_0 \leq T \leq  179.2^\circ $C thermal expansion \\ 
 & 2.486e-4$^\circ $C$^{-1}$  &  6.105e-4 $^\circ $C$^{-1}$ & $T>179.2^\circ $C  thermal expansion \\ 

$\tau^0_B$ & NA & 1e4 s  & BER relaxation time \\

$E_a$ & NA  & 900 J/mol & BER activation energy  \\

$v$ & 0.49  & 0.49 & Poisson ratio \\
$E_{A1}(T)$ &   \multicolumn{3}{l}{Elastic modulus for linear spring element $A_1$ } \\
$E_{A2}(T)$ &  \multicolumn{3}{l}{Elastic modulus for linear spring element $A^s_2$} \\
$E_{A3}(T)$ & \multicolumn{3}{l}{Elastic modulus for linear spring element $A^s_3$ }\\
$\tau^T_2(T)$ & \multicolumn{3}{l}{Thermal relaxation timescale for the element $A^d_2$ }\\
$\tau^T_3(T)$ & \multicolumn{3}{l}{Thermal relaxation timescale for the element $A^d_3$ }\\
\\
\multicolumn{4}{l}{Fiber} \\

$E$ & \multicolumn{2}{c}{22.5GPa}  & Elastic modulus for fiber \\

$\alpha_f$ & \multicolumn{2}{c}{1e-5 $^\circ $C$^{-1}$}  & Thermal expansion coefficient for fiber \\
$v_f$ & \multicolumn{2}{c}{0.15} & Poisson ratio for fiber \citet{hartman1994high}\\

\bottomrule
\end{tabular}

\end{center}

\section{Results and Discussion}  \label{sec:results}

In this section, we first show the TMA experimental results along with a simulation of a neat vitrimer cube. Then, we examine both experimental and simulated behavior for the single fiber composite during the initial curing and cooling stages typically undergone during the manufacture of a fiber composite. We then look at composite behavior over a single thermal cycle that reaches temperatures well into the expected BER regime, representing one potential method of post-cure thermal processing.  Finally, we apply the model to predict the residual stress evolution of an alternative thermal loading history with an even higher peak temperature. 

\subsection{TMA for CTE of the vitrimer}

\begin{figure}[pos=h]
	\centering
		\includegraphics[scale=.25]{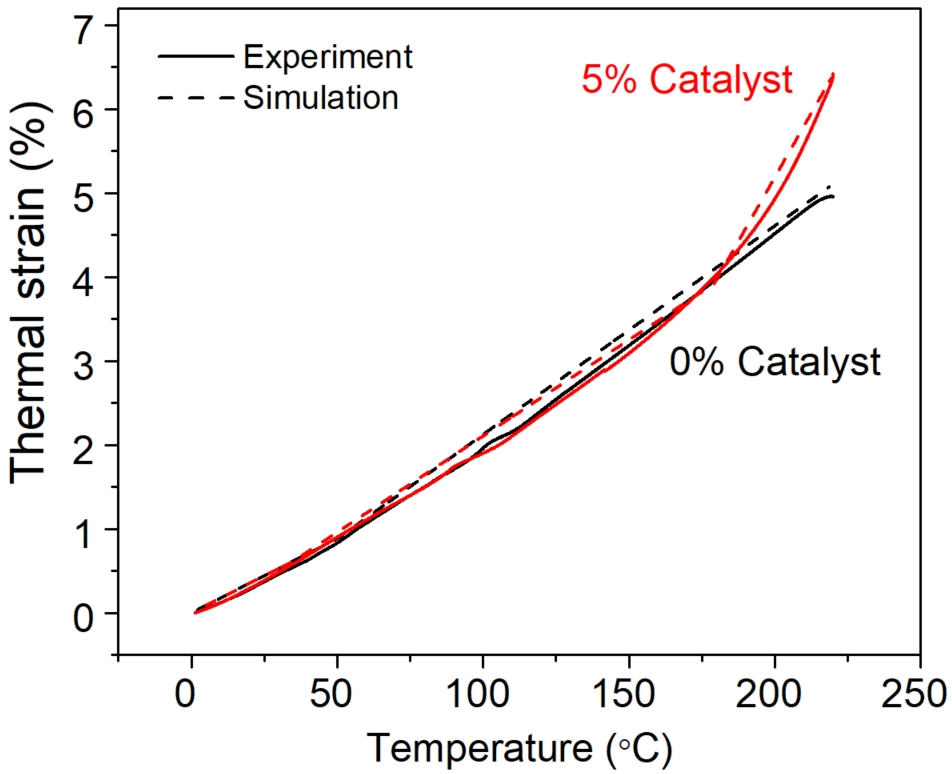}
	      \caption{TMA experiment (solid line) and simulation (dashed line) results for vitrimer with 0$\%$ and 5$\%$ catalyst.}
	\label{fig:tma}
    \end{figure}

The strain evolution of the neat epoxy vitrimer with change in temperature and negligible load applied is shown in Figure \ref{fig:tma}. The experimental data of the vitrimer samples with 0\% (control) and 5$\%$ catalyst are presented. Thermal strain for 0\% catalyst is bilinear with a minor transition around $50 ^\circ $C corresponding to the glass transition temperature. The thermal strain in the 5$\%$ catalyst sample displays three distinct regions: below $35.1 ^\circ $C, between $35.1 ^\circ $C and $179.2 ^\circ $C, and above $179.2 ^\circ $C. $35.1 ^\circ $C and $179.2 ^\circ $C correspond to the glass transition and vitrification temperatures, respectively. We will utilize these glass transition temperatures identified via TMA as the reference temperatures, $T_0$. To verify proper behavior of the material model, the TMA experiment was simulated using Abaqus with the implemented UMAT. A cube representing the neat epoxy vitrimer was modeled under the same boundary conditions and temperature sweep as described in Section \ref{sec:TMA}. Figure \ref{fig:tma} compares the experimental results with the simulation outcomes, demonstrating that the implemented UMAT effectively captures the thermal strain behavior of the vitrimer system for both 0$\%$ and 5$\%$ catalyst concentrations.

\subsection{Curing process of the single fiber composite}

    \begin{figure}[hbt!]
	\centering
		\includegraphics[scale=.35]{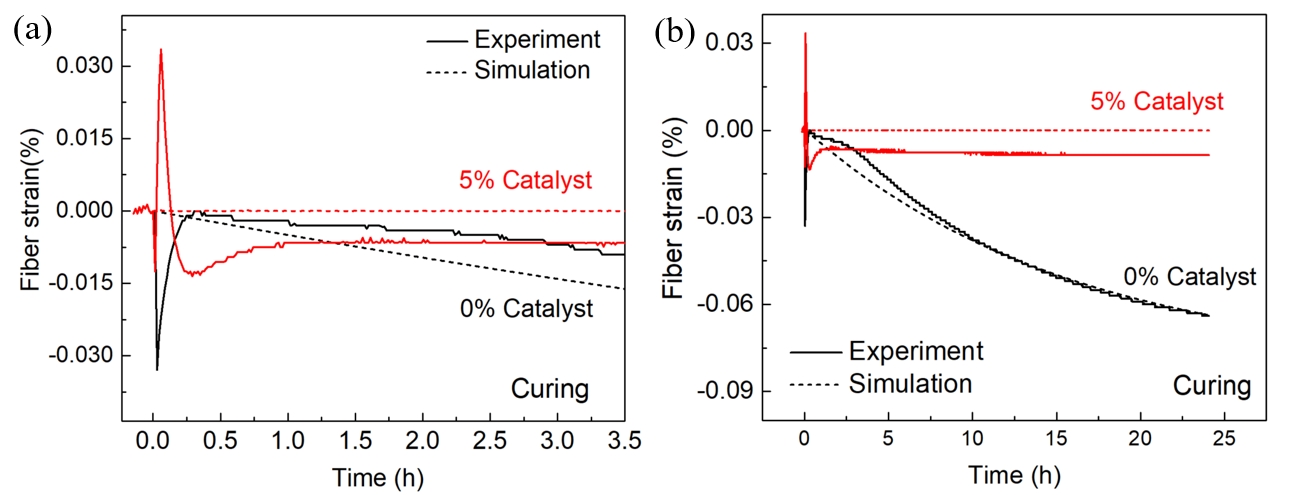}
	      \caption{The simulation and experiment results for fiber strain  with 0$\%$ and 5$\%$ catalyst throughout the curing process The simulation and experimental result for (a) the first 3.5 hours and (b) total cure (24h).}
	\label{fig:cure}
    \end{figure}

During curing, the material transitions from a liquid to a solid structure, and the residual strain induced by this process significantly affects the composite's post-curing properties. The experimental fiber strain for the composite of vitrimers with 0$\%$ (black line) and 5$\%$ (red line) catalyst are shown in Figure \ref{fig:cure}(a) and Figure \ref{fig:cure}(b) for the first 3.5 hours and full 24 hour cure period, respectively. Initially, opening the oven to place the vitrimer into the mold causes a temperature drop and initial fiber strain decrease, while closing the oven allows the temperature to rise, leading to an increase in fiber strain. For the 0$\%$ catalyst vitrimer, the fiber strain decreases roughly linearly the first 3 hours, as seen in Figure \ref{fig:cure}(a), and then follows a long-term exponential decay as shown in Figure \ref{fig:cure}(b). In contrast, the 5$\%$  catalyst vitrimer exhibits a significant spike in fiber strain after the initial phase due to the exothermic reaction triggered by the catalyst. This temperature profile is illustrated in Figure \ref{fig:thermocouple} in Appendix \ref{app:thermocouple}. The temperature rise that results from the exothermic nature of the reaction in the 5$\%$ catalyst vitrimer accelerates the curing process and results in a transient tensile fiber strain from its own thermal expansion. Following this phase, the curing process causes the vitrimer to contract, driving compressive fiber strain.
However, BERs effectively alleviate a significant portion of the cure strain leading to a reduction in fiber strain magnitude. Notably, the magnitude of fiber strain in the vitrimer sample with 5$\%$ catalyst concentration during the curing process is significantly lower than that in the sample with 0$\%$ catalyst, though the cure strain is never fully eliminated. This highlights the effective role of the BERs in reducing residual strain within the material, which we further understand through simulations.

The dotted line in Figure \ref{fig:cure}(a) presents simulation results for vitrimers with 0$\%$ and 5$\%$ catalyst. We fit the fiber strain during cure for the 0$\%$ catalyst to obtain the cure rate, $\tau^{-1}_\phi$, and cure contraction coefficient, $\alpha_C$ parameters. To make the most direct comparison between the 0$\%$ and 5$\%$ catalyst systems, we chose to set the curing model parameters for the 5$\%$ catalyst vitrimer equal to those of the 0$\%$ catalyst vitrimer, even though this underestimates the cure rate for the 5\% catalyst case. For the same reason, we used the applied cure temperature of the oven rather than the temperature within each sample as the input condition for the model. According to Figure \ref{fig:cure}(a), simulation results demonstrate that the fiber strain for the vitrimer with 5$\%$ catalyst is significantly lower than that for the 0$\%$ catalyst, aligning well with experimental observations. The curing process causes vitrimer contraction via element C, while the flow of element B releases this strain. When we adjust the cure parameters specifically for the 5$\%$  catalyst sample the competing dynamics between cure-induced strain and BER flow becomes more obvious, as shown in Figure \ref{fig:cure_para} in Appendix \ref{app:cure_parameter}.

        \begin{figure}[hbt!]
	\centering
		\includegraphics[scale=.25]{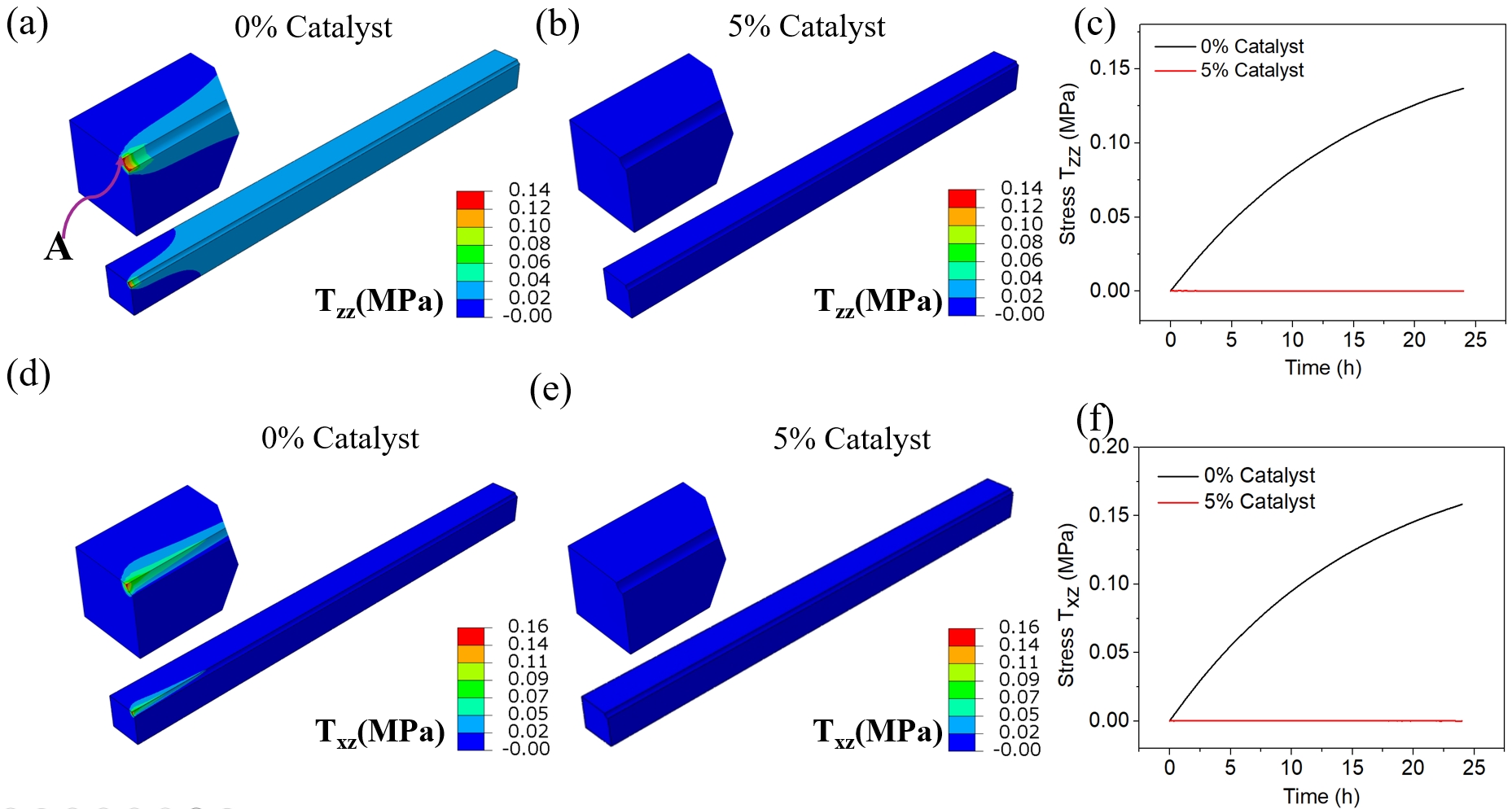}
	      \caption{Cauchy stress component $T_{zz}$ at the end of cure for (a) 0$\%$ catalyst and (b) 5$\%$ catalyst . (c) Cauchy stress component $T_{zz}$ evolution at the point A. Cauchy stress component $T_{xz}$ at the end of cure for  (d) 0$\%$ catalyst and (e) 5$\%$ catalyst. (f) Cauchy stress component  $T_{xz}$ evolution at the point A.}
	\label{fig:curestress}
    \end{figure}

In polymer matrix fiber composites, tensile stress in the matrix, corresponding to compressive stress on the fiber, can lead to fiber failure through mechanisms such as microbuckling, where fibers undergo localized instability under compressive loads \citep{qiao2005explicit,xu2013critical}. Figures \ref{fig:curestress}(a) and (b) compare the true stress along the fiber direction, $T_{zz}$, at the end of the curing process for the simulations with 0\% and 5\% catalyst concentration. In the 0$\%$ catalyst system, the stress distribution is relatively uniform but notable edge effects are present, indicating stress concentrations at the boundaries. In contrast, the true stress $T_{zz}$ in the 5$\%$ catalyst system is significantly reduced, approaching zero. To further analyze the stress evolution during the curing process, we monitored a specific point near the front face of the sample and adjacent to the fiber, labeled as Point A. The results displayed in Figure \ref{fig:curestress}(c) show that while the 0$\%$ catalyst vitrimer exhibits high stress levels during curing, the stress in the 5$\%$ catalyst vitrimer remains minimal. Additionally, shear stress is critical in polymer matrix fiber composites because it governs the load transfer between the polymer matrix and the reinforcing fibers, directly influencing the mechanical performance and structural integrity of the composite \citep{kim1991high,pinho2012material}. Figures \ref{fig:curestress}(d) and (e) illustrate the shear stress $T_{xz}$ for vitrimers with 0$\%$ and 5$\%$ catalysts, respectively. The shear stress $T_{xz}$ is significantly lower in the 5$\%$ catalyst system compared to the 0$\%$ system. Additionally, the evolution of shear stress at Point A during the curing process is shown in Figure \ref{fig:curestress}(f). Unlike the increasing shear stress observed in the 0$\%$ catalyst system, the shear stress in the 5$\%$ catalyst system remains low throughout the cure process. These observations corroborate the experimental finding that the BERs occurring in the 5$\%$ catalyst system effectively reduce residual stresses in the polymer matrix fiber composites.

\subsection{Cooling after curing} \label{sec:cooling}

    \begin{figure}[hbt!]
	\centering
		\includegraphics[scale=.27]{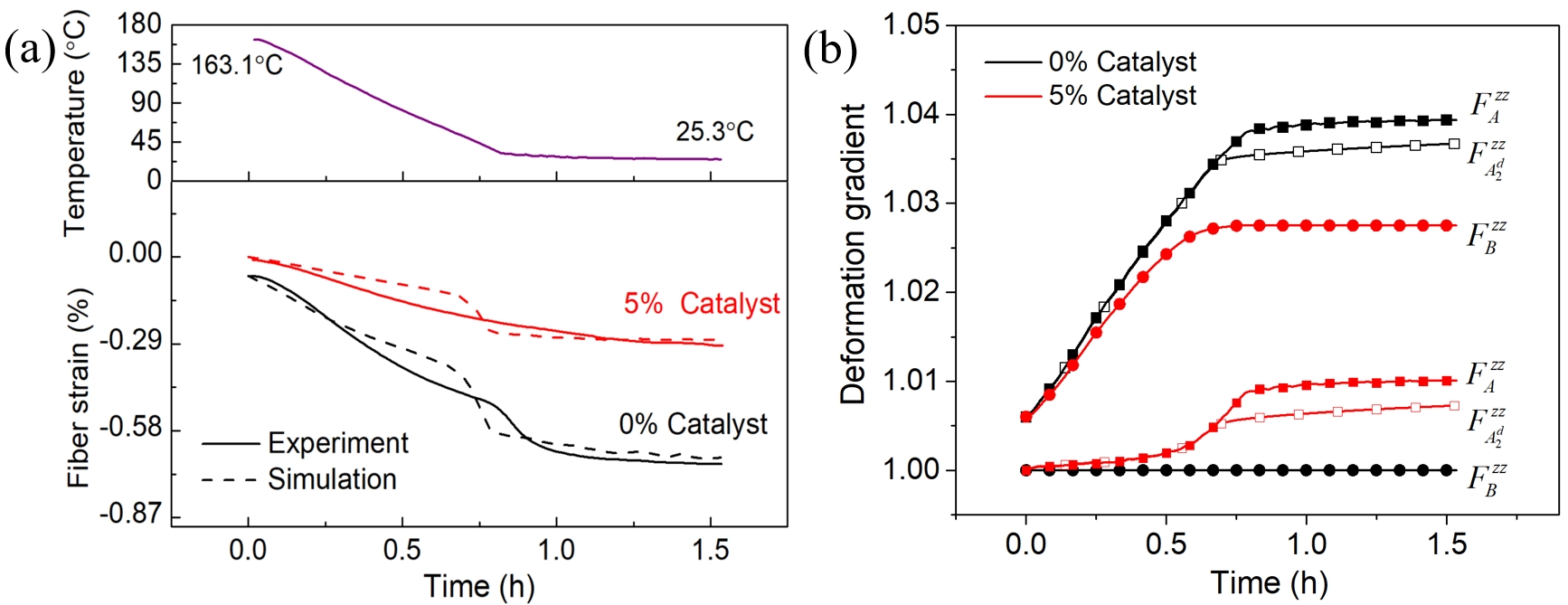}
	      \caption{(a) Temperature profile and the simulation and experimental result for fiber strain in the single fiber composite with 0$\%$ and 5$\%$ catalyst during the post-cure cooling process. (b) A simulation was performed on a cube of neat vitrimer with constraint in the z-direction subject to the same temperature profile as shown in part (a). The deformation gradients along the z-direction for elements labeled $A^d_2$(open squares), $A$(squares) and $B$(circles) are detailed for vitrimer with 0$\%$ and 5$\%$ catalyst during cooling process.}
	\label{fig:cooling}
    \end{figure}

After cure, the composite is cooled to room temperature. Since the vitrimer exhibits a larger CTE than the fiber, it contracts more with the drop in temperature, introducing additional residual interfacial shear stress and compressive stress on the fiber. Figure \ref{fig:cooling}(a) presents both the temperature profile and the corresponding experimental and simulation results for fiber strain. The experimental results for both the 0$\%$ and 5$\%$ catalyst display a similar trend to each other: the fiber strain decreases (increases in magnitude) as the temperature decreases with a slight transition near the glass transition temperature. After the transition, vitrimer with 0$\%$ catalyst shows thermal relaxation behavior during the holding period.  Notably, the fiber strain in the 5$\%$ catalyst vitrimer composite is consistently lower than in the 0$\%$ catalyst sample throughout the cooling process, demonstrating the effective role of BERs in reducing residual strain. The dashed lines in Figure \ref{fig:cooling}(a) depict the simulation results for vitrimers with 0$\%$ and 5$\%$ catalyst. These results indicate that the fiber strain for the vitrimer with 5$\%$ catalyst is significantly lower than that of the 0$\%$ catalyst, aligning with the general trend observed in experimental data. The simulation result for the 0$\%$ catalyst captures the trend in fiber strain especially well, exhibiting both the transition around reference temperature and thermal relaxation behavior.

To more clearly show that the stress relaxation during the post-cure cooling process is indeed arising primarily from BERs, a cube of neat vitrimer was simulated both with 0\% and 5\% catalyst. The model incorporated constraints that prevented displacement in the z-direction and applied symmetry boundary conditions along the xz- and yz-planes capturing a simplified version of the fiber influence. The schematic of the boundary condition is presented in Figure \ref{fig:cube_bc} in Appendix \ref{app:cube_bc}. The simulation results are shown in Figure \ref{fig:cooling}(b). Given the similarity in behavior between dashpots $A^d_2$ and $A^d_3$ during the cooling process, we present only the behavior of element $A^d_2$ for simplicity. For the 0$\%$ catalyst vitrimer, BERs are inactive, so the thermal contraction is accommodated by tensile deformation of element A ($\textbf{F}^{zz}_{A}$). Part of the stress generated in response is alleviated by the flow of viscous stretches $\textbf{F}^{zz}_{A^d_2}$ and $\textbf{F}^{zz}_{A^d_3}$, which barely trail the overall deformation of element A at high temperatures / before 0.6 hours (because of the short time constant) and then fall behind and slowly approach element A following the drop of temperature below the glass transition. In contrast, in the 5$\%$ catalyst vitrimer, BERs are active at high temperatures, so $\textbf{F}^{zz}_{B}$ accommodates much of the tensile deformation imposed by the boundary constraint combined with thermal contraction during cooling, resulting in a significantly reduced $\textbf{F}^{zz}_{A}$ evolution at high temperature. As expected from our choice to keep the 0$\%$ and 5$\%$ catalyst parameters the same for element A, $\textbf{F}^{zz}_{A^d_2}$ trails $\textbf{F}^{zz}_{A}$ for 5$\%$ catalyst analogously to the 0$\%$ catalyst case.

   \begin{figure}[hbt!]
	\centering
		\includegraphics[scale=.25]{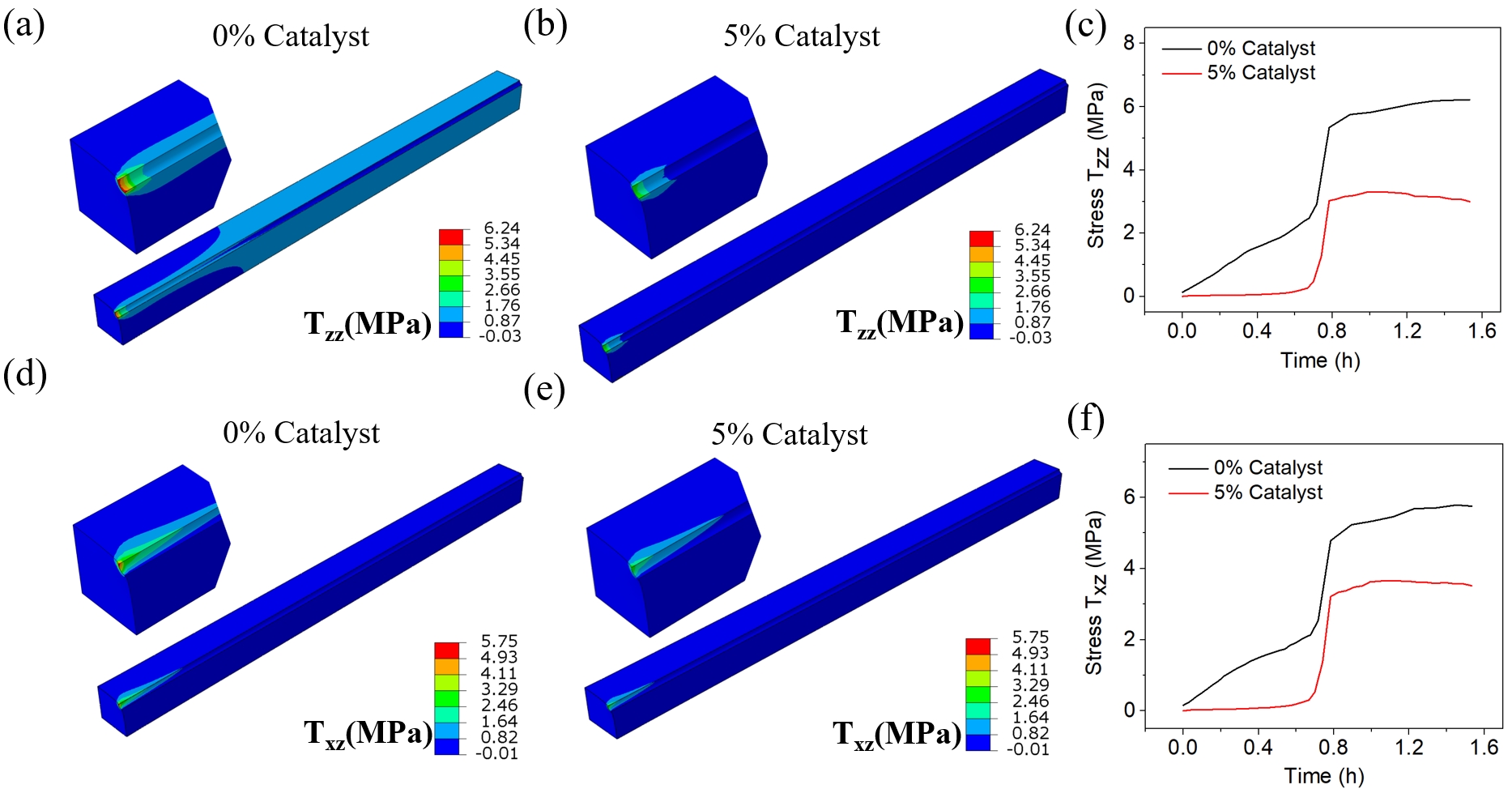}
	      \caption{Cauchy stress component $T_{zz}$ at the end of cooling for (a) 0$\%$ catalyst and (b) 5$\%$ catalyst. (c) Cauchy stress component $T_{zz}$ evolution at point A. Cauchy stress component $T_{xz}$  at the end of cooling for (d) 0$\%$ catalyst and (e) 5$\%$ catalyst. (f) Cauchy stress component $T_{xz}$ evolution at point A.}
	\label{fig:coolingstress}
    \end{figure}

To assess the stress state in the vitrimer matrix during the composite cooling process, we analyzed stress contours for vitrimers with 0$\%$ and 5$\%$ catalyst, as presented in Figure \ref{fig:coolingstress}. Figures \ref{fig:coolingstress}(a) and (b) display the Cauchy stress $T_{zz}$ at the conclusion of the cooling process. As the temperature decreases, the vitrimer matrix, having a higher CTE, attempts to contract more than the fibers. The fibers, due to their high stiffness, resist this contraction, resulting in the matrix experiencing tensile stress. As observed during the curing process, the stress distribution remains relatively uniform in the central region for the 0$\%$ catalyst, but notable edge effects indicate stress concentrations at the boundaries. In contrast, the true stress $T_{zz}$ for the 5$\%$ catalyst is significantly lower. To further explore stress evolution, we again focused on Point A (Figure \ref{fig:coolingstress}(c)). During the cooling process, simulations indicate that stress levels remain high for the 0$\%$ catalyst vitrimer compared with the 5$\%$ catalyst vitrimer. Additionally, shear stress was evaluated. Figures \ref{fig:coolingstress}(d) and (e) show the shear stress $T_{xz}$ for the 0$\%$ and 5$\%$ catalyst, respectively. The shear stress distribution aligns with findings in the literature \citep{zhao2006phase}, showing stress concentration around the corners while remaining relatively uniform across the rest of the region.
 Figure \ref{fig:coolingstress}(f) illustrates the evolution of shear stress at Point A during the cooling process; the 0$\%$ catalyst shows an increase in shear stress, whereas it remains low in the 5$\%$ catalyst system during the cooling process. These results demonstrate that BERs in the 5$\%$ catalyst system effectively mitigate residual stresses in the polymer matrix fiber composites during and after cooling.

\subsection{Thermal cycling}

    \begin{figure}[pos=h]
	\centering
		\includegraphics[scale=.27]{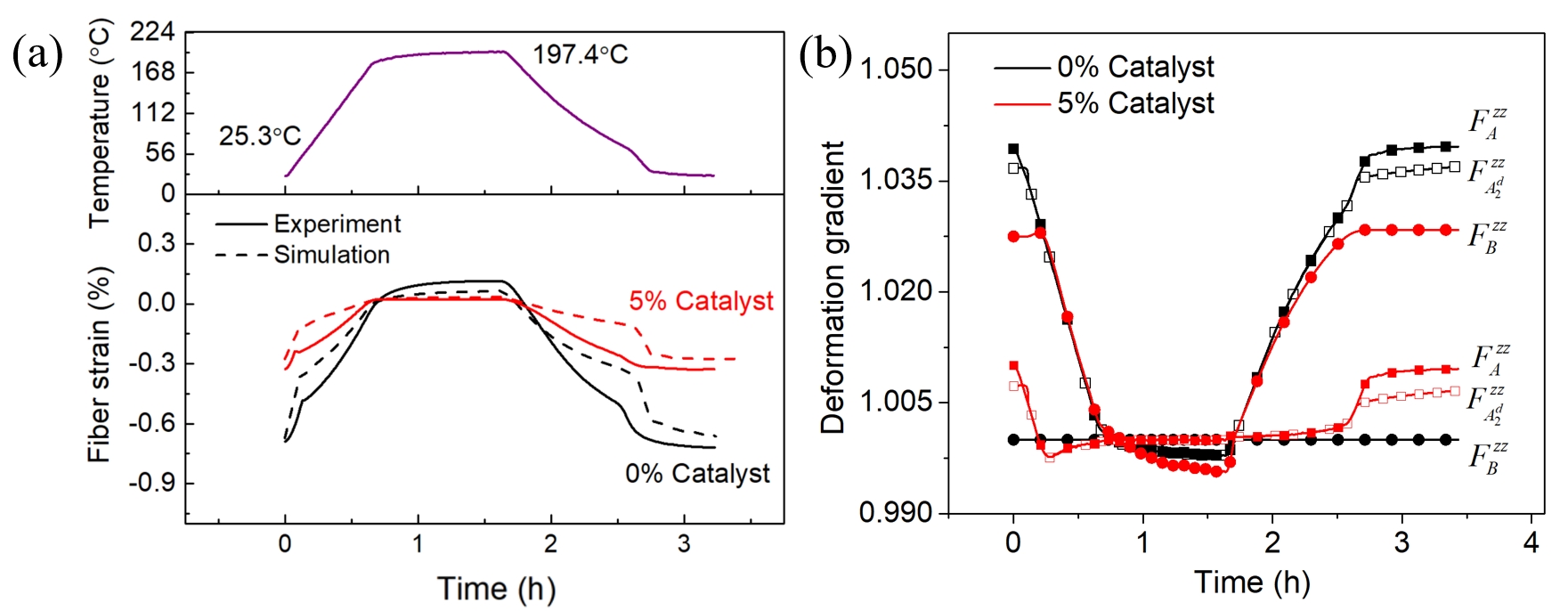}
	      \caption{(a) Temperature profile and the simulation and experimental result for vitrimer with 0$\%$ and 5$\%$ during thermal cycle. (b) The deformation gradients for elements labeled $A^d_2$(open squares), $A$(squares), and $B$(circles) are detailed for vitrimer with 0$\%$ and 5$\%$ catalyst during thermal cycle.}
	\label{fig:thermalcycle}
    \end{figure}

We next examine how a thermal cycle shifts the residual stresses for both the 0$\%$ and 5$\%$ catalyst vitrimer. Figure \ref{fig:thermalcycle}(a) illustrates the temperature profile alongside experimental and simulation results for fiber strain. The temperature cycle begins at 25.3$^\circ$C, increases to 197.4$^\circ$C, and then decreases back to 25$^\circ$C. The experimental data for both the 0$\%$ and 5$\%$ catalyst concentrations exhibit a consistent pattern: fiber strain increases with rising temperatures, displaying a minor transition near the glass transition temperature, and then continues to rise more slowly until the peak temperature is reached. Upon cooling, the fiber strain decreases, undergoes a slight transition, and ultimately stabilizes. Throughout the thermal cycling, the fiber strain in the vitrimer with 5$\%$ catalyst is consistently lower than in the 0$\%$ catalyst because of BERs. The simulation results closely mirror the experimental results, capturing all of the trends including the minor transition around reference temperature. Additionally, the experimental and simulation results for both the 0$\%$ and 5$\%$ catalyst samples indicate that the fiber strain levels before and after the thermal cycle are nearly identical. This suggests that thermal cycling would not be an effective means to reduce residual strain in polymer matrix fiber composites with thermally-triggered BERs in the vitrimeric phase, because any stress released at high temperatures is regained during the cooling process. All experimental data is shown in Figure \ref{fig:exp_all} in Appendix \ref{app:exp_all}.

Figure \ref{fig:thermalcycle}(b) shows the deformation gradient for elements $A$, $A^d_2$ and $B$  for a cube of neat vitrimer subject to no displacement in the z-direction and the same thermal cycle process as the composite (see Appendix \ref{app:cube_bc}). The deformation breakdown results during the thermal cycle are similar to those observed during the cooling process. For the vitrimer with 0$\%$ catalyst, element A compensates for the thermal expansion and contraction, with some of the stress relieved through the viscous elements $\textbf{F}^{zz}_{A^d_2}$ and $\textbf{F}^{zz}_{A^d_3}$ (not shown). In contrast, the vitrimer with 5$\%$ catalyst, accommodates thermal expansion and contraction through a combination of elements A and B, with $\textbf{F}^{zz}_{A}$ varying much less than in the 0$\%$ case.

    \begin{figure}[pos=h]
	\centering
		\includegraphics[scale=.25]{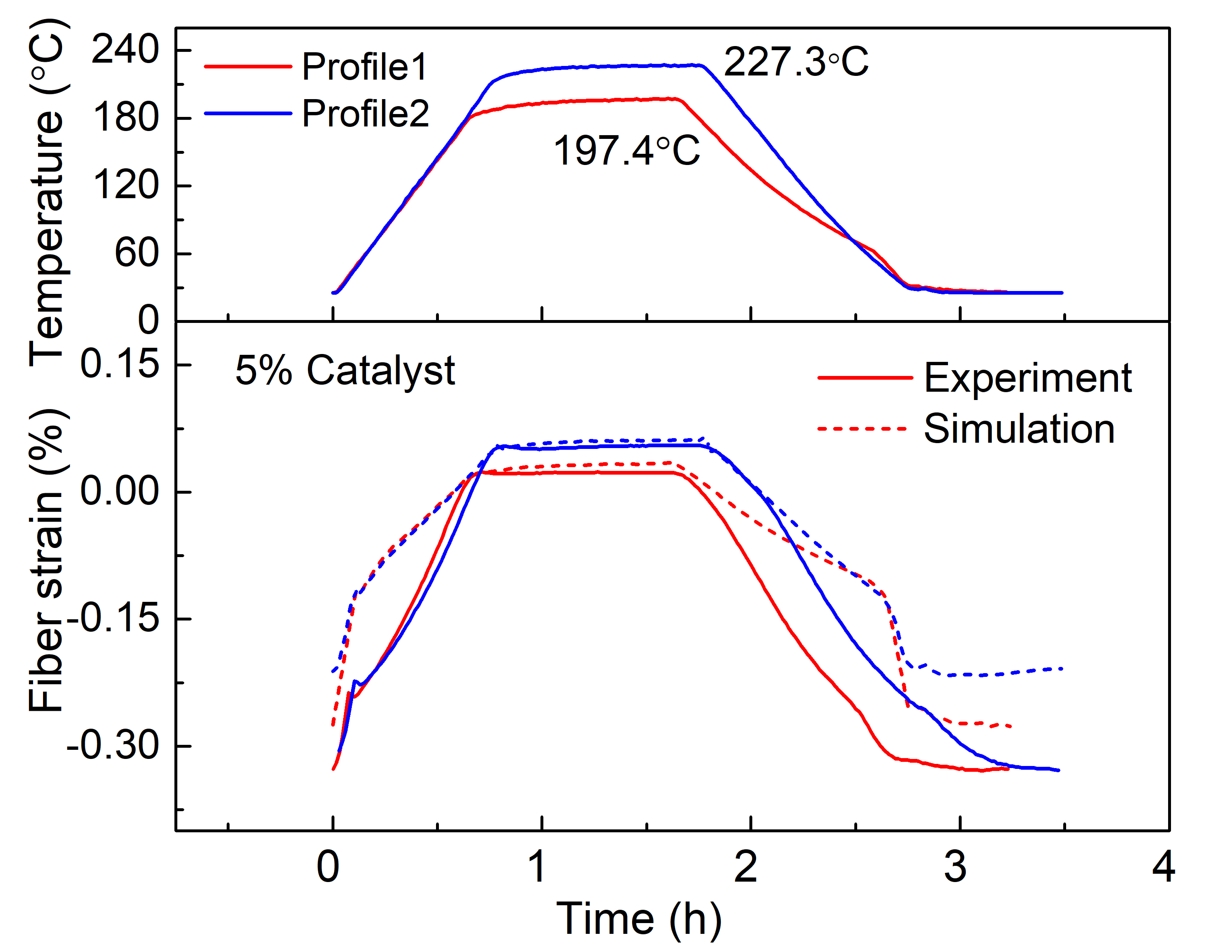}
	      \caption{Temperature profile and the simulation and experimental result for 5$\%$ vitrimer with Profile1 and Profile2. }
	\label{fig:thermalcycle2}
    \end{figure}

To check whether this lack of change with thermal cycling is temperature specific, we carried out further FBG experiments and simulations on the 5$\%$ catalyst vitrimer featuring a thermal cycle to a higher peak temperature of 227.3$^\circ$C (Figure \ref{fig:thermalcycle2}). The fiber strain increases with heating, becomes positive (tensile) at this higher temperature, stays relatively constant, and then decreases. The experimental data and simulation predictions for both thermal cycles demonstrate similar behavior, with the fiber strain remaining virtually unchanged after each cycle. Based on these findings, we conclude that variations in peak temperature during thermal cycles do not significantly impact the residual strain levels after cycling. This insight is crucial for controlling residual stress, guiding the optimization of both the curing and cooling processes to manage stress within the composite effectively.

\section{Conclusion}

In this study, we explore the use of vitrimers enhanced with BERs as a method to alleviate the residual stresses caused by manufacturing processes and differential thermal expansion in polymer matrix fiber composites. We used a single glass fiber in a vitrimer as the representative composite. Experimentally, we formed this representative composite using a Fiber Bragg Grating sensor. The sensor enabled readout of strain on the fiber throughout the cure process and during externally imposed temperature changes. We simulated the representative composite with finite element analysis by devising a 3d finite deformation rheological model for the vitrimer that encompasses cure, thermomechanical relaxation, and BERs, as well as a utilizing a simple linear elastic thermomechanical model for the glass fiber. This simulation partnered with the experiments to provides information on the evolving stress distribution in the vitrimer within the composite. Our experimental and simulation findings indicate that vitrimers containing 5$\%$ catalyst exhibit significantly reduced fiber strain compared to those with 0$\%$ catalyst throughout both the curing and post-curing stages. Overall, our research confirms that vitrimers with BERs can effectively reduce residual stresses, thereby enhancing the structural integrity and longevity of polymer matrix fiber composites.

While this study demonstrates that temperature-activated BERs can mitigate residual stress in polymer matrix fiber composites, several questions remain. First, our experiments focused solely on a system with 5$\%$ catalyst concentration. Exploring other concentrations could potentially enhance our understanding of the effectiveness of BERsacross a broader range of conditions. Second, while thermally triggered BERs prove effective for certain applications, their reliance on temperatures and the lack of precise control over the reaction site are notable limitations. Further research into alternative activation methods for BERs is warranted. In particular, light-triggered BERs present a promising option due to their precise control and ease of activation, which could lead to more effective reduction of residual stresses.

\section{Acknowledgments}
This work was supported by the United States National Air Force Office of Scientific Research under Contract No. FA9550-22-1-0030. This work made use of the Cornell Center for Materials Research Facilities supported by the National Science Foundation under Award Number DMR-1719875. This work made use of the Analytical $\&$ Diagnostics Laboratory at Binghamton University’s S3IP Center of Excellence. We thank Dr. S Chen at Cornell University for the valuable discussions.  We thank Dr. D Nepel at the Air Force Research Laboratory (AFRL) for the guidance and support in the synthesis of vitrimer materials.

\appendix

\counterwithin{figure}{section}

\section{Parameter identification} \label{app:para_ident}

\subsection{Temperature dependent  Young’s modulus and timescale  identification} \label{app:compress}

  \begin{figure}[h]
	\centering
		\includegraphics[scale=.27]{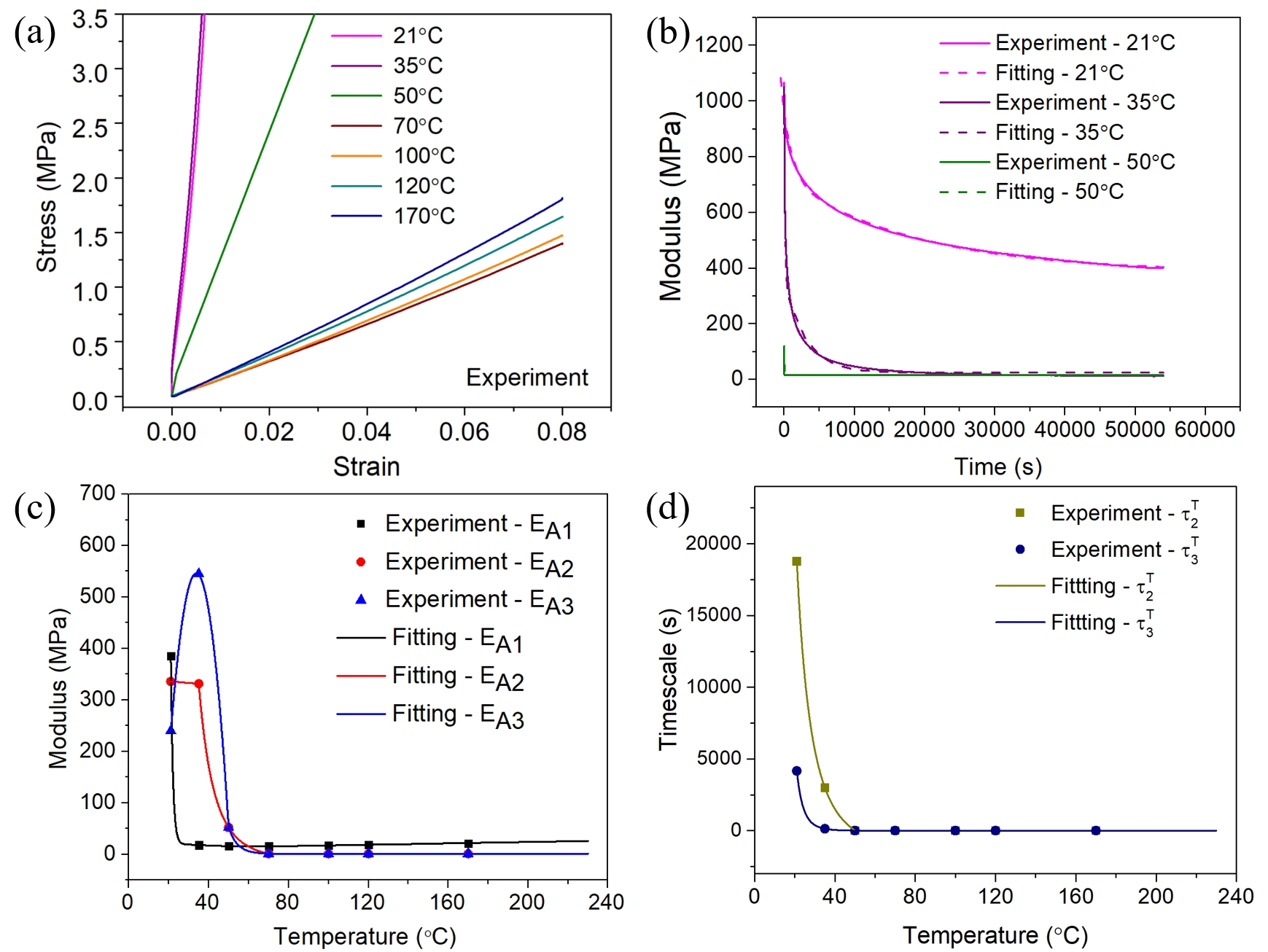}
	      \caption{(a) Compression test results at 21°C, 35°C, 50°C, 70°C, 100°C, 120°C, and 170°C; (b) Compression relaxation experiment and fitting results at 21°C, 35°C, and 50°C; (c) Values of $E_{A1}$, $E_{A2}$, and $E_{A3}$ from compression relaxation experiments (dots) and the fitting function (solid line) used in the UMAT; (d) Values of $\tau^T_2$ and $\tau^T_3$ from compression relaxation experiments (dots) and the fitting function (solid line) used in the UMAT.}
	\label{fig:compress}
    \end{figure}

Compression relaxation experiments were performed to characterize the temperature dependent modulus and relaxation for each branch. The modulus can expressed as

\begin{align}
    &  E(T) = E_{A1} (T) +   E_{A2}(T) \exp \left(-\frac{t}{\tau^T_2}\right) +  E_{A3}(T)\exp\left(-\frac{t}{\tau^T_3}\right) \label{eq:compression_rex}
\end{align}
\noindent where $ E_{A1}$, $ E_{A2}$, and $ E_{A3}$ are the elastic moduli for the branch springs $A_1$, $A^s_2$ and $A^s_3$, respectively. $\tau^T_2$ and $\tau^T_3$ are the thermal relaxation timescale for the dashpots $A^d_2$ and $A^d_3$, respectively. For temperatures of 21$^\circ$C, 35$^\circ$C, and 50$^\circ$C, the fitting parameters were obtained using Equation \ref{eq:compression_rex} with a nonlinear least squares method, as listed in Table \ref{tab:compress}. The fitting results are presented in Figure \ref{fig:compress}(b). At higher temperatures (70$^\circ$C, 100$^\circ$C, 120$^\circ$C, and 170$^\circ$C), vitrimer relaxation occurs too rapidly to be accurately captured by the Zwick testing machine. To maintain consistency with the fitting results at 50$^\circ$C, we set the timescale to 10 seconds for these higher temperatures, which induces sufficiently fast relaxation over the 10$^3$ s timescales investigated. For the elastic modulus, the non-equilibrium branch moduli, $E_{A2}$ and $E_{A3}$, are minimal at high temperatures and were set arbitrarily to 1 MPa (a low value) for each branch. The modulus $E_{A1}$ was calculated based on the initial modulus obtained from the compression test shown in Figure \ref{fig:compress}(a). After obtaining these above parameters at each temperature, we interpolated between temperatures using a smooth function to determine the modulus and relaxation time scale for each branch at arbitrary temperatures for implementation into the UMAT. The smooth functions for moduli $E_{A1}$, $E_{A2}$, and $E_{A3}$ are shown as solid lines in Figure \ref{fig:compress}(c), while those for timescales $\tau^T_2$ and $\tau^T_3$ are shown as solid lines in Figure \ref{fig:compress}(d).

\begin{center}
\captionof{table}{Modulus and timescale parameters} \label{tab:compress}
\begin{tabular}{llllll} 
\toprule
{Temperature($^\circ$C)} & {$E_{A1}$(MPa)} & {$E_{A2}$(MPa)} & {$E_{A3}$(MPa)}&{$\tau^T_2$(s)} &  {$\tau^T_3$(s)}  \\ 
\midrule
21 &  385.1 & 335.9& 239.9 & 18770 & 4165\\ 
35 &  18 & 331.5& 545 &3000 & 10 \\
50 &  15.6 & 52&52 & 10 & 10\\
70 &  15.4 & 1 & 1 & 10 & 10\\
100 &  17.1 & 1 & 1 & 10 & 10\\
120 &  18.5 & 1 & 1 & 10 & 10\\
170 &  20.9 & 1 & 1 & 10 & 10\\
\bottomrule
\end{tabular}
\end{center}

\subsection{Fitting parameters from Fiber Bragg Grating Experiments} \label{app:fbg_ident}

In the curing process described in Section \ref{sec:element1}, we use two key parameters: the cure contraction coefficient $\alpha_c $ , and the characteristic cure rate $\tau_\phi^{-1}$. By fitting the fiber strain curve for the $0\%$ catalyst sample during curing, as shown in Figure \ref{fig:cure}(b), we obtained $\alpha_c = 0.77$ and the   $\tau_\phi^{-1} = 1.75\times 10^{-5}$ s$^{-1}$. For a direct comparison between the 0$\%$ and 5$\%$ catalyst systems, we set the curing model parameters for the 5$\%$ catalyst vitrimer equal to those determined for the 0$\%$ catalyst vitrimer.

In Section \ref{sec:element3}, the viscous BER flow associated with element B is characterized by two parameters related to the BER relaxation timescales: the reference relaxation time $\tau_B^0$ and   the BER activation energy $E_a$. By fitting the fiber strain curve for the 5$\%$ catalyst sample during the cooling process, shown in Figure \ref{fig:cooling}(a), we obtained $\tau_B^0 = 1\times 10^4$ s and $E_a = 900$ J/mol.

\subsection{Coefficient of thermal expansion} \label{app:TMA_ident}

The TMA experimental result is shown in Figure \ref{fig:tma}. Thermal strain for 0\% catalyst is bilinear with a minor transition around $50 ^\circ$C, which we set as the reference temperature $T_0$. The CTE, $\alpha_{nT}$, above and below $T_0$ are taken as the slopes of these two linear regions. For temperatures below $T_0$, $\alpha_{nT} = 1.716 \times 10^{-4}$ $^\circ$C$^{-1}$ and for temperatures above $T_0$, $\alpha_{nT} = 2.486 \times 10^{-4}$ $^\circ$C$^{-1}$. The thermal strain in the 5$\%$ catalyst sample displays three distinct regions: below $35.1 ^\circ$C, between $35.1 ^\circ$C and $179.2 ^\circ$C, and above $179.2 ^\circ$C. The reference temperature $T_0$ for 5$\%$ catalyst was set to $35.1 ^\circ$C. Above $179.2 ^\circ$C the strain escalates dramatically due to the activation of BER, resulting in a nonlinear response. Therefore, $179.2 ^\circ$C was taken as the vitrification temperature, $T_v$, above which BER become prominent. The CTE for the 5$\%$ catalyst vitrimer is calculated based on the slopes of these three regions. For temperatures below $T_0$, $\alpha_{nT} = 1.765 \times 10^{-4}$ $^\circ$C$^{-1}$. For temperatures above $T_0$ and below $T_v$, $\alpha_{nT} = 2.291 \times 10^{-4}$ $^\circ$C$^{-1}$. For temperatures above $T_v$, $\alpha_{nT} = 6.105\times 10^{-4}$ $^\circ$C$^{-1}$.

\subsection{Fiber Bragg Grating Sensor Fiber Experimental Characterization} \label{app:fiber_exp}

 \begin{figure}[h]
	\centering
		\includegraphics[scale=.27]{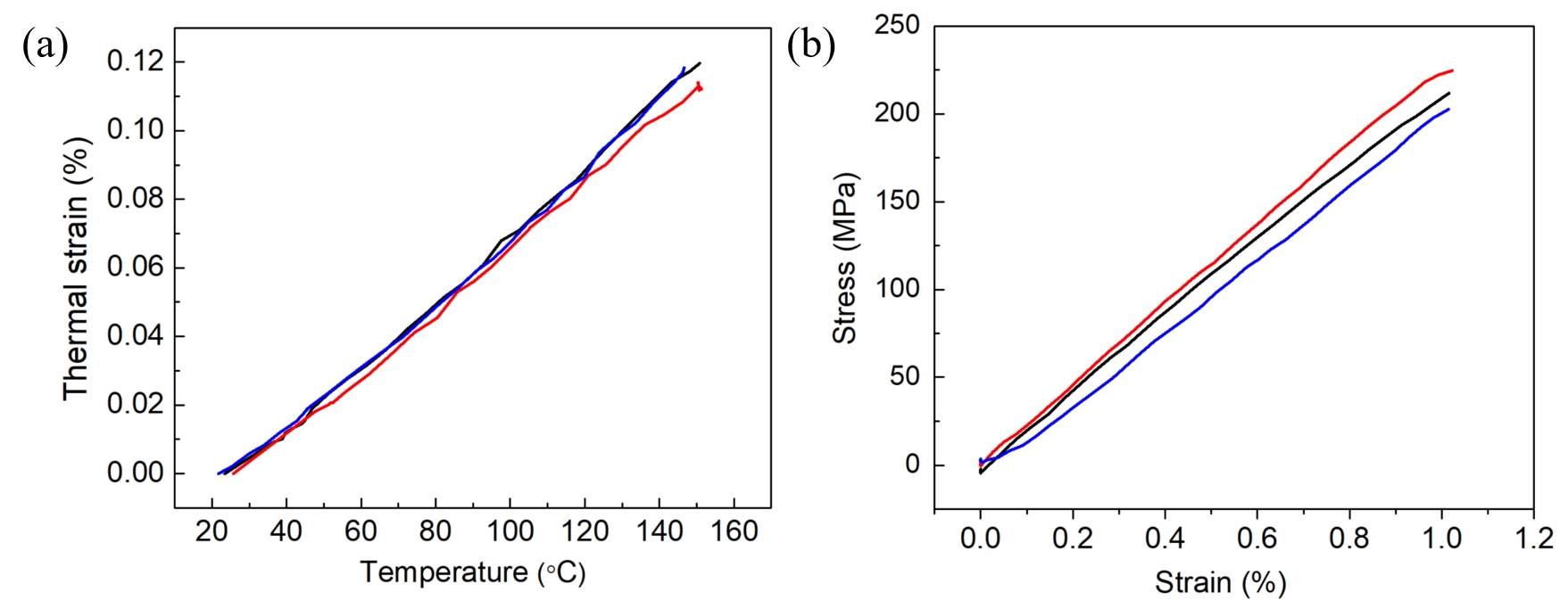}
	      \caption{(a) The experiment of the CTE of the glass fiber. (b) Uniaxial tension test for the glass fiber.}
	\label{fig:fiber_prop}
\end{figure}

To determine the coefficient of thermal expansion (CTE) of the Fiber Bragg Grating fiber, we employed a Fiber Bragg Grating (FBG) sensor system. The glass fiber was placed inside an oven with the temperature set to increase from 20$ ^\circ$C to 150$ ^\circ$C at a rate of 5$ ^\circ$C per minute. A thermocouple was also positioned near the glass fiber to accurately measure its temperature. The results, shown in Figure \ref{fig:fiber_prop}(a), indicate a linear relationship between the thermal strain of the glass fiber and the temperature. The CTE of the glass fiber was calculated from the slope of the thermal strain curve, resulting in a value of $\alpha_f = 1 \times {10^{-5}}$ $ ^\circ$C$^{-1}$.\\

To determine the elastic modulus of the Fiber Bragg Grating, which is glass with a thin polymer coating, a uniaxial tension test was carried out. The grips used to hold the fiber were created using a 3D printer. The glass fiber was then carefully positioned on the grip and secured using epoxy glue. The assembly was left undisturbed for one day to ensure that the epoxy fully cured and the fiber was securely bonded to the grip. The tension test was performed using a Zwick/Roell Z010 universal testing machine equipped with a 10kN load cell, operating at room temperature. The test was conducted under crosshead displacement control with constant engineering strain rates to ensure accuracy. Figure \ref{fig:fiber_prop}(b) displays the stress-strain relationship for the glass fiber. From the slope of this curve, the elastic modulus of the glass fiber was calculated to be $22.5$ GPa.

\section{Implicit integration approach} \label{app:scheme}

The model developed in the paper was implemented into a UMAT subroutine in ABAQUS. We implement an iterative implicit integration scheme adapted from \citep{ma2014photoviscoplastic} as follows:

\begin{enumerate}

\item Update the applied deformation gradient $\textbf{F}$ as imposed by external boundary conditions.  

\item Compute the shift factors ($\alpha_B$) and relaxation timescales ($\tau_B$).

\item Compute the isotropic cure strain ($\epsilon_c$) and thermal strain ($\epsilon_T$). The deformation gradient ($\textbf{F}_C$) for the element C is then taken as: $\textbf{F}_C = \exp\left(\epsilon_c\right)\exp\left(\epsilon_T\right) \textbf{I}$.

\item Use $\textbf{F}_C$ to compute the pre-relaxation deformation gradients governing the intermediate stress responses as $\textbf{F}_{A}= \textbf{F}_B^{-1} \textbf{F}_C^{-1} \textbf{F}$, $\textbf{F}_{A^s_2}=\textbf{F}_{A^d_2}^{-1} \textbf{F}_{A}$ and $\textbf{F}_{A^s_3}=\textbf{F}_{A^d_3}^{-1} \textbf{F}_{A}$, where $\textbf{F}_B=\textbf{I}$, $\textbf{F}_{A^d_2}=\textbf{I}$ and  $\textbf{F}_{A^d_3}=\textbf{I}$ prior to relaxation.  

\item Using the pre-relaxation deformation gradients $\textbf{F}_{A}$, $\textbf{F}_{A^s_2}$ and $\textbf{F}_{A^s_3}$, compute the corresponding true stresses of each branch  $\textbf{T}_{A1}$, $\textbf{T}_{A2}$ and $\textbf{T}_{A3}$, respectively.

\item Using $\textbf{T}_{A2}$, compute viscous stretch $\textbf{F}_{A^d_2}$ through Eq (17) to (20).

\item Using $\textbf{T}_{A3}$, compute viscous stretch $\textbf{F}_{A^d_3}$ through Eq (21) to (24).

\item Given updated values of $\textbf{F}_{A^d_2}$ and $\textbf{F}_{A^d_3}$ (while still that maintaining $\textbf{F}_B=\textbf{I}$), recompute the components of the deformation gradient as $\textbf{F}_{A}= \textbf{F}_B^{-1} \textbf{F}_C^{-1} \textbf{F}$, $\textbf{F}_{A^s_2}=\textbf{F}_{A^d_2}^{-1} \textbf{F}_{A}$ and $\textbf{F}_{A^s_3}=\textbf{F}_{A^d_3}^{-1} \textbf{F}_{A}$.

\item Using $\textbf{F}_{A}$, $\textbf{F}_{A^s_2}$ and $\textbf{F}_{A^s_3}$ , recompute the true stress of each branch $\textbf{T}_{A1}$, $\textbf{T}_{A2}$, $\textbf{T}_{A3}$ and the total stress $\textbf{T}$ through Eq (8) to (13).

\item Using the total stress $\textbf{T}$, compute the relaxation deformation $\textbf{F}_{B}$ due to BER-enabled flow using Eq (25) to (30).

\item Recompute the components of internal deformation a final time as $\textbf{F}_{A}= \textbf{F}_B^{-1} \textbf{F}_C^{-1} \textbf{F}$, $\textbf{F}_{A^s_2}=\textbf{F}_{A^d_2}^{-1} \textbf{F}_{A}$ and $\textbf{F}_{A^s_3}=\textbf{F}_{A^d_3}^{-1} \textbf{F}_{A}$, where now $\bm{F}_{A^d_2}$, $\bm{F}_{A^d_3}$ and $\textbf{F}_B$ are updated to account for relaxation.

\item Using $\textbf{F}_{A^s_2}$ and $\textbf{F}_{A^s_3}$, recompute the true stress of each branch $\textbf{T}_{A1}$, $\textbf{T}_{A2}$, $\textbf{T}_{A3}$ and the total stress $\textbf{T}$, and then update the applied strain using the Newton-Raphson method.

\item Iterate through steps 4-12 until the applied traction boundary conditions are satisfied.

\end{enumerate}

\section{Thermocouple Data} \label{app:thermocouple}
   \begin{figure}[h]
	\centering
		\includegraphics[scale=.26]{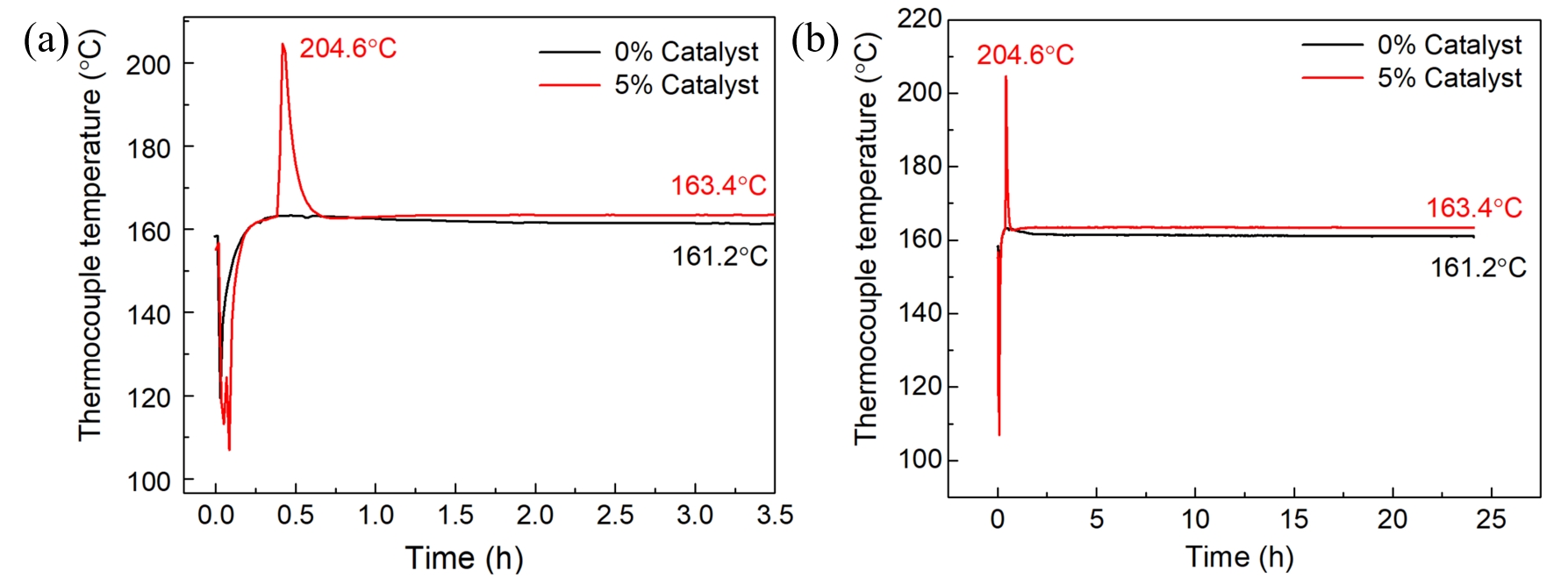}
	      \caption{A representative thermocouple temperature embedded into the vitrimer with 0$\%$ and 5$\%$ catalyst for (a) the first 3.5 hours and (b) the total cure period (24h).}
	\label{fig:thermocouple}
    \end{figure}

To accurately monitor the temperature change, thermocouples were inserted into both concentrations of vitrimer samples to record the temperature during the curing process. The temperature profile, shown in Figure \ref{fig:thermocouple}(a), indicates a slight increase to approximately 163$^\circ$C before stabilizing at 161.2$^\circ$C for the 0$\%$ catalyst sample (black curve). For the 5$\%$ catalyst sample (red curve), there is an initial sharp temperature rise to 204.6$^\circ$C, followed by a decrease and stabilization at 163.4$^\circ$C. Ultimately, the temperature stabilizes at approximately 161.2$^\circ$C for the 0$\%$ catalyst and 163.4$^\circ$C for the 5$\%$ catalyst, as illustrated in Figure \ref{fig:thermocouple}(b).

\section{Cure parameter study} \label{app:cure_parameter}

\begin{figure}[h]
	\centering
		\includegraphics[scale=.25]{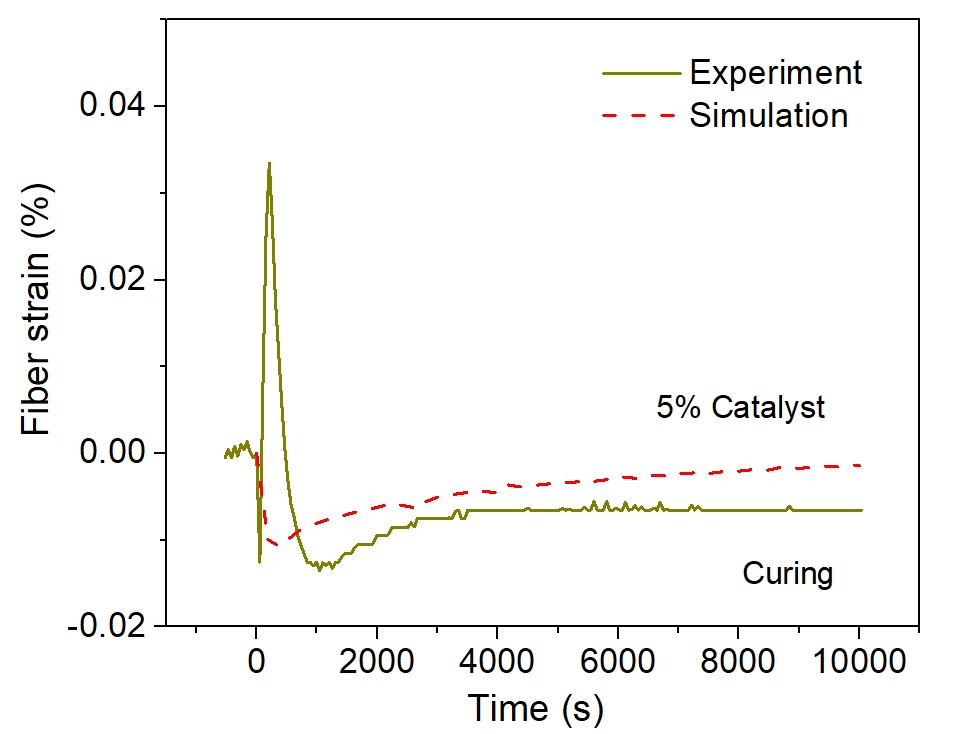}
	      \caption{Fiber strain with different cure parameters.}
	\label{fig:cure_para}
    \end{figure}

Catalysts play a pivotal role in the curing process by accelerating the chemical reactions that transform monomers into a cross-linked polymer network. This acceleration is achieved by lowering the activation energy required for the reaction, leading to faster curing times and potentially altering the material's final properties \citep{lukin2020platinum,kim2024triazabicyclodecene}. As observed from the experiments in Figure \ref{fig:cure}(a), the relatively fast, catalyzed and exothermic curing process for the 5\% catalyst vitrimer leads to an increase in temperature within the vitrimer that further accelerates the cure rate. This increase in local temperature influences the cure rate $\tau_\phi^{-1}$ and cure contraction coefficient $\alpha_c$ for the cure model in Section \ref{sec:element1}. Here, we modified the cure parameters to capture a representative interaction between cure-induced strain and BER flow if this exothermic acceleration were included in the model. Specifically, the cure rate $\tau_\phi^{-1}$ was adjusted to $1.75\times 10^{-4}$, and the cure contraction coefficient $\alpha_c$ was set to 0.2. We simulated the curing behavior over approximately 10,000 seconds. This simulation highlights the competing dynamics between cure-induced strain and BER flow. As shown in Figure \ref{fig:cure_para}, the curing process causes the material to contract, initially dominating the fiber strain. Meanwhile, the BER flow leads to a decrease in strain magnitude. These opposing mechanisms create an interaction between the BER flow of element B and the cure-induced strain. The BERs effectively mitigate a significant portion of the cure strain until the two effects reach a balance, resulting in an initial increase in fiber strain magnitude, followed by a decrease, and eventual stabilization.
    
\section{The boundary condition for the constrained cube simulations} \label{app:cube_bc}

\begin{figure}[hbt!]
	\centering
		\includegraphics[scale=.4]{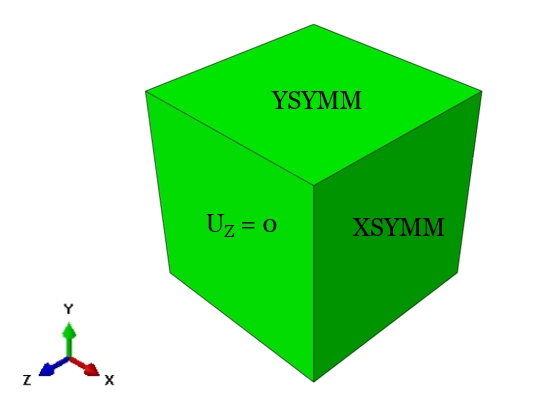}
	      \caption{The boundary condition for the cube simulations are as follows: the left surface is constrained in Z-direction with $U_z = 0$ and the opposite (right) surface is similarly constrained. The top surface and front surface are subjected to YSYMM and XSYMM respectively.}
	\label{fig:cube_bc}
    \end{figure}

Figure \ref{fig:cube_bc} shows the boundary condition for the constrained cube simulations.

\section{Vitrimer Experimental Data} \label{app:exp_all}

   \begin{figure}[hbt!]
	\centering
		\includegraphics[scale=.4]{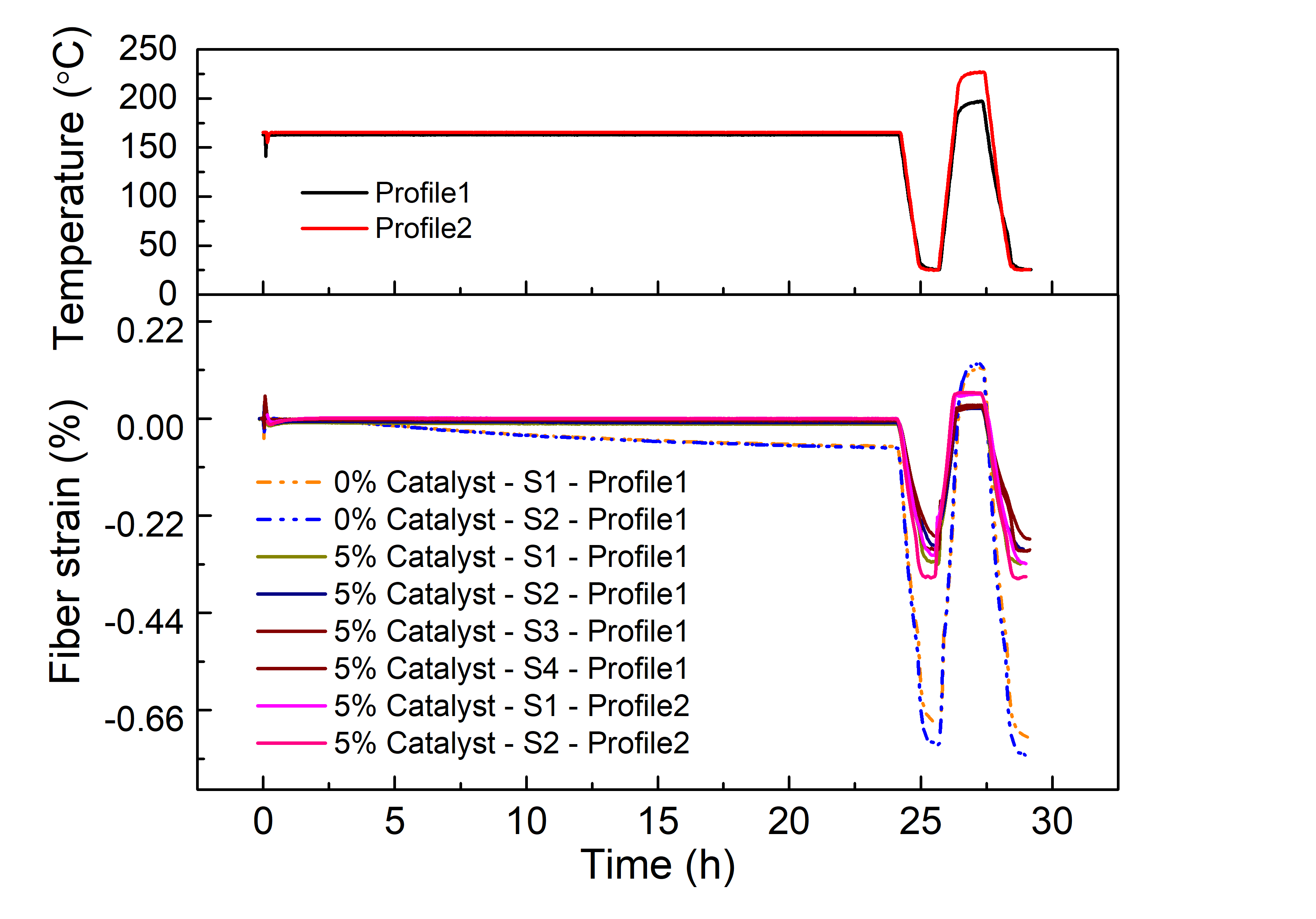}
	      \caption{Temperature profile and experimental results for vitrimers with  0$\%$ and 5$\%$ catalyst concentrations.}
	\label{fig:exp_all}
    \end{figure}

A Fiber Bragg Grating (FBG) sensor system was placed into vitrimer system to measure the fiber strain of the fiber composite. The top of Figure \ref{fig:exp_all} details two different thermal cycles: Profile 1 starts at 163.1$^\circ$C for 24 hours, cools down to 25.3$^\circ$C, heats up to 197.4$^\circ$C, and then returns to 25.3$^\circ$C. Profile 2 follows a similar pattern to Profile 1 but with a peak temperature of 227.3$^\circ$C instead of 197.4$^\circ$C. The bottom of Figure \ref{fig:exp_all} displays the experimental results for fiber strain under varying catalyst concentrations. For the 0$\%$ catalyst concentration in Profile 1, experiments were conducted on two samples, labeled S1 and S2. In the case of the 5$\%$ catalyst concentration with Profile 1, four samples were tested, ranging from S1 to S4. For the 5$\%$ catalyst concentration with Profile 2, two samples, S1 and S2, were assessed.

\printcredits

\bibliographystyle{cas-model2-names}
\bibliography{Reference}

\begin{thebibliography}{64}
\expandafter\ifx\csname natexlab\endcsname\relax\def\natexlab#1{#1}\fi
\providecommand{\url}[1]{\texttt{#1}}
\providecommand{\href}[2]{#2}
\providecommand{\path}[1]{#1}
\providecommand{\DOIprefix}{doi:}
\providecommand{\ArXivprefix}{arXiv:}
\providecommand{\URLprefix}{URL: }
\providecommand{\Pubmedprefix}{pmid:}
\providecommand{\doi}[1]{\href{http://dx.doi.org/#1}{\path{#1}}}
\providecommand{\Pubmed}[1]{\href{pmid:#1}{\path{#1}}}
\providecommand{\bibinfo}[2]{#2}
\ifx\xfnm\relax \def\xfnm[#1]{\unskip,\space#1}\fi
\bibitem[{Adolf et~al.(1998)Adolf, Martin, Chambers, Burchett and
  Guess}]{adolf1998stresses}
\bibinfo{author}{Adolf, D.B.}, \bibinfo{author}{Martin, J.E.},
  \bibinfo{author}{Chambers, R.S.}, \bibinfo{author}{Burchett, S.N.},
  \bibinfo{author}{Guess, T.R.}, \bibinfo{year}{1998}.
\newblock \bibinfo{title}{Stresses during thermoset cure}.
\newblock \bibinfo{journal}{Journal of materials research}
  \bibinfo{volume}{13}, \bibinfo{pages}{530--550}.
\bibitem[{An et~al.(2022)An, Shi, Jin, Zhao and Wang}]{an2022chain}
\bibinfo{author}{An, L.}, \bibinfo{author}{Shi, Q.}, \bibinfo{author}{Jin, C.},
  \bibinfo{author}{Zhao, W.}, \bibinfo{author}{Wang, T.}, \bibinfo{year}{2022}.
\newblock \bibinfo{title}{Chain diffusion based framework for modeling the
  welding of vitrimers}.
\newblock \bibinfo{journal}{Journal of the Mechanics and Physics of Solids}
  \bibinfo{volume}{164}, \bibinfo{pages}{104883}.
\bibitem[{Baran et~al.(2017)Baran, Cinar, Ersoy, Akkerman and
  Hattel}]{baran2017review}
\bibinfo{author}{Baran, I.}, \bibinfo{author}{Cinar, K.},
  \bibinfo{author}{Ersoy, N.}, \bibinfo{author}{Akkerman, R.},
  \bibinfo{author}{Hattel, J.H.}, \bibinfo{year}{2017}.
\newblock \bibinfo{title}{A review on the mechanical modeling of composite
  manufacturing processes}.
\newblock \bibinfo{journal}{Archives of computational methods in engineering}
  \bibinfo{volume}{24}, \bibinfo{pages}{365--395}.
\bibitem[{Burns et~al.(2016)Burns, Mouritz, Pook and
  Feih}]{burns2016strengthening}
\bibinfo{author}{Burns, L.}, \bibinfo{author}{Mouritz, A.},
  \bibinfo{author}{Pook, D.}, \bibinfo{author}{Feih, S.}, \bibinfo{year}{2016}.
\newblock \bibinfo{title}{Strengthening of composite t-joints using novel ply
  design approaches}.
\newblock \bibinfo{journal}{Composites Part B: Engineering}
  \bibinfo{volume}{88}, \bibinfo{pages}{73--84}.
\bibitem[{Chava et~al.(2022)Chava, Namilae and Al-Haik}]{chava2022residual}
\bibinfo{author}{Chava, S.}, \bibinfo{author}{Namilae, S.},
  \bibinfo{author}{Al-Haik, M.}, \bibinfo{year}{2022}.
\newblock \bibinfo{title}{Residual stress reduction during composite
  manufacturing through cure modification: In situ analysis}.
\newblock \bibinfo{journal}{Journal of Composite Materials}
  \bibinfo{volume}{56}, \bibinfo{pages}{975--988}.
\bibitem[{Chen and Zhang(2019)}]{chen2019improved}
\bibinfo{author}{Chen, W.}, \bibinfo{author}{Zhang, D.}, \bibinfo{year}{2019}.
\newblock \bibinfo{title}{Improved prediction of residual stress induced
  warpage in thermoset composites using a multiscale thermo-viscoelastic
  processing model}.
\newblock \bibinfo{journal}{Composites Part A: Applied Science and
  Manufacturing} \bibinfo{volume}{126}, \bibinfo{pages}{105575}.
\bibitem[{Danzi et~al.(2019)Danzi, Fanteria, Panettieri and
  Mancino}]{danzi2019numerical}
\bibinfo{author}{Danzi, F.}, \bibinfo{author}{Fanteria, D.},
  \bibinfo{author}{Panettieri, E.}, \bibinfo{author}{Mancino, M.},
  \bibinfo{year}{2019}.
\newblock \bibinfo{title}{A numerical micro-mechanical study on damage induced
  by the curing process in carbon/epoxy unidirectional material}.
\newblock \bibinfo{journal}{Composite Structures} \bibinfo{volume}{210},
  \bibinfo{pages}{755--766}.
\bibitem[{Denissen et~al.(2016)Denissen, Winne and
  Du~Prez}]{denissen2016vitrimers}
\bibinfo{author}{Denissen, W.}, \bibinfo{author}{Winne, J.M.},
  \bibinfo{author}{Du~Prez, F.E.}, \bibinfo{year}{2016}.
\newblock \bibinfo{title}{Vitrimers: permanent organic networks with glass-like
  fluidity}.
\newblock \bibinfo{journal}{Chemical science} \bibinfo{volume}{7},
  \bibinfo{pages}{30--38}.
\bibitem[{Duflou et~al.(2012)Duflou, Deng, Van~Acker and
  Dewulf}]{duflou2012fiber}
\bibinfo{author}{Duflou, J.R.}, \bibinfo{author}{Deng, Y.},
  \bibinfo{author}{Van~Acker, K.}, \bibinfo{author}{Dewulf, W.},
  \bibinfo{year}{2012}.
\newblock \bibinfo{title}{Do fiber-reinforced polymer composites provide
  environmentally benign alternatives? a life-cycle-assessment-based study}.
\newblock \bibinfo{journal}{Mrs Bulletin} \bibinfo{volume}{37},
  \bibinfo{pages}{374--382}.
\bibitem[{D’mello et~al.(2016)D’mello, Maiar{\`u} and Waas}]{d2016virtual}
\bibinfo{author}{D’mello, R.}, \bibinfo{author}{Maiar{\`u}, M.},
  \bibinfo{author}{Waas, A.}, \bibinfo{year}{2016}.
\newblock \bibinfo{title}{Virtual manufacturing of composite aerostructures}.
\newblock \bibinfo{journal}{The Aeronautical Journal} \bibinfo{volume}{120},
  \bibinfo{pages}{61--81}.
\bibitem[{D’Mello and Waas(2017)}]{d2017virtual}
\bibinfo{author}{D’Mello, R.J.}, \bibinfo{author}{Waas, A.M.},
  \bibinfo{year}{2017}.
\newblock \bibinfo{title}{Virtual curing of textile polymer matrix composites}.
\newblock \bibinfo{journal}{Composite Structures} \bibinfo{volume}{178},
  \bibinfo{pages}{455--466}.
\bibitem[{Fazal and Fancey(2014)}]{fazal2014uhmwpe}
\bibinfo{author}{Fazal, A.}, \bibinfo{author}{Fancey, K.S.},
  \bibinfo{year}{2014}.
\newblock \bibinfo{title}{Uhmwpe fibre-based composites: Prestress-induced
  enhancement of impact properties}.
\newblock \bibinfo{journal}{Composites Part B: Engineering}
  \bibinfo{volume}{66}, \bibinfo{pages}{1--6}.
\bibitem[{Fernlund et~al.(2002)Fernlund, Rahman, Courdji, Bresslauer,
  Poursartip, Willden and Nelson}]{fernlund2002experimental}
\bibinfo{author}{Fernlund, G.}, \bibinfo{author}{Rahman, N.},
  \bibinfo{author}{Courdji, R.}, \bibinfo{author}{Bresslauer, M.},
  \bibinfo{author}{Poursartip, A.}, \bibinfo{author}{Willden, K.},
  \bibinfo{author}{Nelson, K.}, \bibinfo{year}{2002}.
\newblock \bibinfo{title}{Experimental and numerical study of the effect of
  cure cycle, tool surface, geometry, and lay-up on the dimensional fidelity of
  autoclave-processed composite parts}.
\newblock \bibinfo{journal}{Composites part A: applied science and
  manufacturing} \bibinfo{volume}{33}, \bibinfo{pages}{341--351}.
\bibitem[{Fish et~al.(2021)Fish, Wagner and Keten}]{fish2021mesoscopic}
\bibinfo{author}{Fish, J.}, \bibinfo{author}{Wagner, G.J.},
  \bibinfo{author}{Keten, S.}, \bibinfo{year}{2021}.
\newblock \bibinfo{title}{Mesoscopic and multiscale modelling in materials}.
\newblock \bibinfo{journal}{Nature materials} \bibinfo{volume}{20},
  \bibinfo{pages}{774--786}.
\bibitem[{Ghasemi et~al.(2015)Ghasemi, Mohammadi and
  Mohandes}]{ghasemi2015role}
\bibinfo{author}{Ghasemi, A.}, \bibinfo{author}{Mohammadi, M.},
  \bibinfo{author}{Mohandes, M.}, \bibinfo{year}{2015}.
\newblock \bibinfo{title}{The role of carbon nanofibers on thermo-mechanical
  properties of polymer matrix composites and their effect on reduction of
  residual stresses}.
\newblock \bibinfo{journal}{Composites Part B: Engineering}
  \bibinfo{volume}{77}, \bibinfo{pages}{519--527}.
\bibitem[{Hartman et~al.(1994)Hartman, Greenwood and Miller}]{hartman1994high}
\bibinfo{author}{Hartman, D.}, \bibinfo{author}{Greenwood, M.E.},
  \bibinfo{author}{Miller, D.M.}, \bibinfo{year}{1994}.
\newblock \bibinfo{title}{High strength glass fibers}.
\newblock \bibinfo{journal}{Moving Forward With 50 Years of Leadership in
  Advanced Materials.} \bibinfo{volume}{39}, \bibinfo{pages}{521--533}.
\bibitem[{He et~al.(2019)He, Ge, Qi, Gao, Chen, Liang and
  Fang}]{he2019multiscale}
\bibinfo{author}{He, C.}, \bibinfo{author}{Ge, J.}, \bibinfo{author}{Qi, D.},
  \bibinfo{author}{Gao, J.}, \bibinfo{author}{Chen, Y.},
  \bibinfo{author}{Liang, J.}, \bibinfo{author}{Fang, D.},
  \bibinfo{year}{2019}.
\newblock \bibinfo{title}{A multiscale elasto-plastic damage model for the
  nonlinear behavior of 3d braided composites}.
\newblock \bibinfo{journal}{Composites Science and Technology}
  \bibinfo{volume}{171}, \bibinfo{pages}{21--33}.
\bibitem[{Herrera-Franco and Valadez-Gonzalez(2005)}]{herrera2005study}
\bibinfo{author}{Herrera-Franco, P.}, \bibinfo{author}{Valadez-Gonzalez, A.},
  \bibinfo{year}{2005}.
\newblock \bibinfo{title}{A study of the mechanical properties of short
  natural-fiber reinforced composites}.
\newblock \bibinfo{journal}{Composites Part B: Engineering}
  \bibinfo{volume}{36}, \bibinfo{pages}{597--608}.
\bibitem[{Hu et~al.(2018)Hu, Cao, Pavier, Zhong, Zu, Liu and
  Li}]{hu2018investigation}
\bibinfo{author}{Hu, H.}, \bibinfo{author}{Cao, D.}, \bibinfo{author}{Pavier,
  M.}, \bibinfo{author}{Zhong, Y.}, \bibinfo{author}{Zu, L.},
  \bibinfo{author}{Liu, L.}, \bibinfo{author}{Li, S.}, \bibinfo{year}{2018}.
\newblock \bibinfo{title}{Investigation of non-uniform gelation effects on
  residual stresses of thick laminates based on tailed fbg sensor}.
\newblock \bibinfo{journal}{Composite Structures} \bibinfo{volume}{202},
  \bibinfo{pages}{1361--1372}.
\bibitem[{Hubbard et~al.(2021)Hubbard, Ren, Konkolewicz, Sarvestani, Picu,
  Kedziora, Roy, Varshney and Nepal}]{hubbard2021vitrimer}
\bibinfo{author}{Hubbard, A.M.}, \bibinfo{author}{Ren, Y.},
  \bibinfo{author}{Konkolewicz, D.}, \bibinfo{author}{Sarvestani, A.},
  \bibinfo{author}{Picu, C.R.}, \bibinfo{author}{Kedziora, G.S.},
  \bibinfo{author}{Roy, A.}, \bibinfo{author}{Varshney, V.},
  \bibinfo{author}{Nepal, D.}, \bibinfo{year}{2021}.
\newblock \bibinfo{title}{Vitrimer transition temperature identification:
  coupling various thermomechanical methodologies}.
\newblock \bibinfo{journal}{ACS Applied Polymer Materials} \bibinfo{volume}{3},
  \bibinfo{pages}{1756--1766}.
\bibitem[{Hubbard et~al.(2022)Hubbard, Ren, Picu, Sarvestani, Konkolewicz, Roy,
  Varshney and Nepal}]{hubbard2022creep}
\bibinfo{author}{Hubbard, A.M.}, \bibinfo{author}{Ren, Y.},
  \bibinfo{author}{Picu, C.R.}, \bibinfo{author}{Sarvestani, A.},
  \bibinfo{author}{Konkolewicz, D.}, \bibinfo{author}{Roy, A.K.},
  \bibinfo{author}{Varshney, V.}, \bibinfo{author}{Nepal, D.},
  \bibinfo{year}{2022}.
\newblock \bibinfo{title}{Creep mechanics of epoxy vitrimer materials}.
\newblock \bibinfo{journal}{ACS Applied Polymer Materials} \bibinfo{volume}{4},
  \bibinfo{pages}{4254--4263}.
\bibitem[{Hui et~al.(2021)Hui, Xu and Zhang}]{hui2021integrated}
\bibinfo{author}{Hui, X.}, \bibinfo{author}{Xu, Y.}, \bibinfo{author}{Zhang,
  W.}, \bibinfo{year}{2021}.
\newblock \bibinfo{title}{An integrated modeling of the curing process and
  transverse tensile damage of unidirectional cfrp composites}.
\newblock \bibinfo{journal}{Composite Structures} \bibinfo{volume}{263},
  \bibinfo{pages}{113681}.
\bibitem[{Hyer and White(2009)}]{hyer2009stress}
\bibinfo{author}{Hyer, M.W.}, \bibinfo{author}{White, S.R.},
  \bibinfo{year}{2009}.
\newblock \bibinfo{title}{Stress analysis of fiber-reinforced composite
  materials}.
\newblock \bibinfo{publisher}{DEStech Publications, Inc}.
\bibitem[{Kang et~al.(2014)Kang, Kim, Lim and Bolander}]{kang2014modeling}
\bibinfo{author}{Kang, J.}, \bibinfo{author}{Kim, K.}, \bibinfo{author}{Lim,
  Y.M.}, \bibinfo{author}{Bolander, J.E.}, \bibinfo{year}{2014}.
\newblock \bibinfo{title}{Modeling of fiber-reinforced cement composites:
  Discrete representation of fiber pullout}.
\newblock \bibinfo{journal}{International Journal of Solids and Structures}
  \bibinfo{volume}{51}, \bibinfo{pages}{1970--1979}.
\bibitem[{Kim et~al.(2024)Kim, Lee and Lee}]{kim2024triazabicyclodecene}
\bibinfo{author}{Kim, J.G.}, \bibinfo{author}{Lee, G.S.}, \bibinfo{author}{Lee,
  A.}, \bibinfo{year}{2024}.
\newblock \bibinfo{title}{Triazabicyclodecene: A versatile catalyst for polymer
  synthesis}.
\newblock \bibinfo{journal}{Journal of Polymer Science} \bibinfo{volume}{62},
  \bibinfo{pages}{42--91}.
\bibitem[{Kim and Mai(1991)}]{kim1991high}
\bibinfo{author}{Kim, J.K.}, \bibinfo{author}{Mai, Y.W.}, \bibinfo{year}{1991}.
\newblock \bibinfo{title}{High strength, high fracture toughness fibre
  composites with interface control—a review}.
\newblock \bibinfo{journal}{Composites Science and Technology}
  \bibinfo{volume}{41}, \bibinfo{pages}{333--378}.
\bibitem[{Kim et~al.(2012)Kim, Murayama, Kageyama, Uzawa and
  Kanai}]{kim2012study}
\bibinfo{author}{Kim, S.S.}, \bibinfo{author}{Murayama, H.},
  \bibinfo{author}{Kageyama, K.}, \bibinfo{author}{Uzawa, K.},
  \bibinfo{author}{Kanai, M.}, \bibinfo{year}{2012}.
\newblock \bibinfo{title}{Study on the curing process for carbon/epoxy
  composites to reduce thermal residual stress}.
\newblock \bibinfo{journal}{Composites Part A: Applied Science and
  Manufacturing} \bibinfo{volume}{43}, \bibinfo{pages}{1197--1202}.
\bibitem[{Liu et~al.(2018)Liu, Hao, Zhang, Yang, Wang, Han, Li, Xin and
  Zhang}]{liu2018self}
\bibinfo{author}{Liu, T.}, \bibinfo{author}{Hao, C.}, \bibinfo{author}{Zhang,
  S.}, \bibinfo{author}{Yang, X.}, \bibinfo{author}{Wang, L.},
  \bibinfo{author}{Han, J.}, \bibinfo{author}{Li, Y.}, \bibinfo{author}{Xin,
  J.}, \bibinfo{author}{Zhang, J.}, \bibinfo{year}{2018}.
\newblock \bibinfo{title}{A self-healable high glass transition temperature
  bioepoxy material based on vitrimer chemistry}.
\newblock \bibinfo{journal}{Macromolecules} \bibinfo{volume}{51},
  \bibinfo{pages}{5577--5585}.
\bibitem[{Liu et~al.(2013)Liu, Greene, Chen, Dikin and
  Liu}]{liu2013computational}
\bibinfo{author}{Liu, Y.}, \bibinfo{author}{Greene, M.S.},
  \bibinfo{author}{Chen, W.}, \bibinfo{author}{Dikin, D.A.},
  \bibinfo{author}{Liu, W.K.}, \bibinfo{year}{2013}.
\newblock \bibinfo{title}{Computational microstructure characterization and
  reconstruction for stochastic multiscale material design}.
\newblock \bibinfo{journal}{Computer-Aided Design} \bibinfo{volume}{45},
  \bibinfo{pages}{65--76}.
\bibitem[{Liu et~al.(2021)Liu, Liu, Li, Weng and Zeng}]{liu2021biobased}
\bibinfo{author}{Liu, Y.Y.}, \bibinfo{author}{Liu, G.L.}, \bibinfo{author}{Li,
  Y.D.}, \bibinfo{author}{Weng, Y.}, \bibinfo{author}{Zeng, J.B.},
  \bibinfo{year}{2021}.
\newblock \bibinfo{title}{Biobased high-performance epoxy vitrimer with uv
  shielding for recyclable carbon fiber reinforced composites}.
\newblock \bibinfo{journal}{ACS Sustainable Chemistry \& Engineering}
  \bibinfo{volume}{9}, \bibinfo{pages}{4638--4647}.
\bibitem[{LLorca et~al.(2011)LLorca, Gonz{\'a}lez, Molina-Aldaregu{\'\i}a,
  Segurado, Seltzer, Sket, Rodr{\'\i}guez, S{\'a}daba, Mu{\~n}oz and
  Canal}]{llorca2011multiscale}
\bibinfo{author}{LLorca, J.}, \bibinfo{author}{Gonz{\'a}lez, C.},
  \bibinfo{author}{Molina-Aldaregu{\'\i}a, J.M.}, \bibinfo{author}{Segurado,
  J.}, \bibinfo{author}{Seltzer, R.}, \bibinfo{author}{Sket, F.},
  \bibinfo{author}{Rodr{\'\i}guez, M.}, \bibinfo{author}{S{\'a}daba, S.},
  \bibinfo{author}{Mu{\~n}oz, R.}, \bibinfo{author}{Canal, L.P.},
  \bibinfo{year}{2011}.
\newblock \bibinfo{title}{Multiscale modeling of composite materials: a roadmap
  towards virtual testing}.
\newblock \bibinfo{journal}{Advanced materials} \bibinfo{volume}{23},
  \bibinfo{pages}{5130--5147}.
\bibitem[{Lukin et~al.(2020)Lukin, Kuchkaev, Sukhov, Bekmukhamedov and
  Yakhvarov}]{lukin2020platinum}
\bibinfo{author}{Lukin, R.Y.}, \bibinfo{author}{Kuchkaev, A.M.},
  \bibinfo{author}{Sukhov, A.V.}, \bibinfo{author}{Bekmukhamedov, G.E.},
  \bibinfo{author}{Yakhvarov, D.G.}, \bibinfo{year}{2020}.
\newblock \bibinfo{title}{Platinum-catalyzed hydrosilylation in polymer
  chemistry}.
\newblock \bibinfo{journal}{Polymers} \bibinfo{volume}{12},
  \bibinfo{pages}{2174}.
\bibitem[{Ma et~al.(2014)Ma, Mu, Bowman, Sun, Dunn, Qi and
  Fang}]{ma2014photoviscoplastic}
\bibinfo{author}{Ma, J.}, \bibinfo{author}{Mu, X.}, \bibinfo{author}{Bowman,
  C.N.}, \bibinfo{author}{Sun, Y.}, \bibinfo{author}{Dunn, M.L.},
  \bibinfo{author}{Qi, H.J.}, \bibinfo{author}{Fang, D.}, \bibinfo{year}{2014}.
\newblock \bibinfo{title}{A photoviscoplastic model for photoactivated covalent
  adaptive networks}.
\newblock \bibinfo{journal}{Journal of the Mechanics and Physics of Solids}
  \bibinfo{volume}{70}, \bibinfo{pages}{84--103}.
\bibitem[{Mao et~al.(2019)Mao, Chen, Hou, Qi and Yu}]{mao2019viscoelastic}
\bibinfo{author}{Mao, Y.}, \bibinfo{author}{Chen, F.}, \bibinfo{author}{Hou,
  S.}, \bibinfo{author}{Qi, H.J.}, \bibinfo{author}{Yu, K.},
  \bibinfo{year}{2019}.
\newblock \bibinfo{title}{A viscoelastic model for hydrothermally activated
  malleable covalent network polymer and its application in shape memory
  analysis}.
\newblock \bibinfo{journal}{Journal of the Mechanics and Physics of Solids}
  \bibinfo{volume}{127}, \bibinfo{pages}{239--265}.
\bibitem[{Mesogitis et~al.(2014)Mesogitis, Skordos and
  Long}]{mesogitis2014uncertainty}
\bibinfo{author}{Mesogitis, T.S.}, \bibinfo{author}{Skordos, A.A.},
  \bibinfo{author}{Long, A.C.}, \bibinfo{year}{2014}.
\newblock \bibinfo{title}{Uncertainty in the manufacturing of fibrous
  thermosetting composites: A review}.
\newblock \bibinfo{journal}{Composites Part A: Applied Science and
  Manufacturing} \bibinfo{volume}{57}, \bibinfo{pages}{67--75}.
\bibitem[{Mohamed et~al.(2020)Mohamed, Brahma, Ning and
  Pillay}]{mohamed2020development}
\bibinfo{author}{Mohamed, M.}, \bibinfo{author}{Brahma, S.},
  \bibinfo{author}{Ning, H.}, \bibinfo{author}{Pillay, S.},
  \bibinfo{year}{2020}.
\newblock \bibinfo{title}{Development of beneficial residual stresses in glass
  fiber epoxy composites through fiber prestressing}.
\newblock \bibinfo{journal}{Journal of Reinforced Plastics and Composites}
  \bibinfo{volume}{39}, \bibinfo{pages}{487--498}.
\bibitem[{Montarnal et~al.(2011)Montarnal, Capelot, Tournilhac and
  Leibler}]{montarnal2011silica}
\bibinfo{author}{Montarnal, D.}, \bibinfo{author}{Capelot, M.},
  \bibinfo{author}{Tournilhac, F.}, \bibinfo{author}{Leibler, L.},
  \bibinfo{year}{2011}.
\newblock \bibinfo{title}{Silica-like malleable materials from permanent
  organic networks}.
\newblock \bibinfo{journal}{Science} \bibinfo{volume}{334},
  \bibinfo{pages}{965--968}.
\bibitem[{Mostafa et~al.(2017)Mostafa, Ismarrubie, Sapuan and
  Sultan}]{mostafa2017fibre}
\bibinfo{author}{Mostafa, N.H.}, \bibinfo{author}{Ismarrubie, Z.},
  \bibinfo{author}{Sapuan, S.}, \bibinfo{author}{Sultan, M.},
  \bibinfo{year}{2017}.
\newblock \bibinfo{title}{Fibre prestressed polymer-matrix composites: a
  review}.
\newblock \bibinfo{journal}{Journal of Composite Materials}
  \bibinfo{volume}{51}, \bibinfo{pages}{39--66}.
\bibitem[{Nagaraj et~al.(2024)Nagaraj, Shah, Sabato and
  Maiaru}]{nagaraj2024validation}
\bibinfo{author}{Nagaraj, M.}, \bibinfo{author}{Shah, S.},
  \bibinfo{author}{Sabato, A.}, \bibinfo{author}{Maiaru, M.},
  \bibinfo{year}{2024}.
\newblock \bibinfo{title}{Validation and verification of a novel higher-order
  fe framework for process modeling of thermoset composites}.
\newblock \bibinfo{journal}{Composites Part B: Engineering}
  \bibinfo{volume}{279}, \bibinfo{pages}{111447}.
\bibitem[{Pinho et~al.(2012)Pinho, Darvizeh, Robinson, Schuecker and
  Camanho}]{pinho2012material}
\bibinfo{author}{Pinho, S.}, \bibinfo{author}{Darvizeh, R.},
  \bibinfo{author}{Robinson, P.}, \bibinfo{author}{Schuecker, C.},
  \bibinfo{author}{Camanho, P.}, \bibinfo{year}{2012}.
\newblock \bibinfo{title}{Material and structural response of polymer-matrix
  fibre-reinforced composites}.
\newblock \bibinfo{journal}{Journal of Composite Materials}
  \bibinfo{volume}{46}, \bibinfo{pages}{2313--2341}.
\bibitem[{Prashanth et~al.(2017)Prashanth, Subbaya, Nithin and
  Sachhidananda}]{prashanth2017fiber}
\bibinfo{author}{Prashanth, S.}, \bibinfo{author}{Subbaya, K.},
  \bibinfo{author}{Nithin, K.}, \bibinfo{author}{Sachhidananda, S.},
  \bibinfo{year}{2017}.
\newblock \bibinfo{title}{Fiber reinforced composites-a review}.
\newblock \bibinfo{journal}{J. Mater. Sci. Eng} \bibinfo{volume}{6},
  \bibinfo{pages}{2--6}.
\bibitem[{Qiao and Shan(2005)}]{qiao2005explicit}
\bibinfo{author}{Qiao, P.}, \bibinfo{author}{Shan, L.}, \bibinfo{year}{2005}.
\newblock \bibinfo{title}{Explicit local buckling analysis and design of
  fiber--reinforced plastic composite structural shapes}.
\newblock \bibinfo{journal}{Composite Structures} \bibinfo{volume}{70},
  \bibinfo{pages}{468--483}.
\bibitem[{Qureshi(2022)}]{qureshi2022review}
\bibinfo{author}{Qureshi, J.}, \bibinfo{year}{2022}.
\newblock \bibinfo{title}{A review of fibre reinforced polymer structures}.
\newblock \bibinfo{journal}{Fibers} \bibinfo{volume}{10}, \bibinfo{pages}{27}.
\bibitem[{R{\"o}ttger et~al.(2017)R{\"o}ttger, Domenech, van Der~Weegen,
  Breuillac, Nicola{\"y} and Leibler}]{rottger2017high}
\bibinfo{author}{R{\"o}ttger, M.}, \bibinfo{author}{Domenech, T.},
  \bibinfo{author}{van Der~Weegen, R.}, \bibinfo{author}{Breuillac, A.},
  \bibinfo{author}{Nicola{\"y}, R.}, \bibinfo{author}{Leibler, L.},
  \bibinfo{year}{2017}.
\newblock \bibinfo{title}{High-performance vitrimers from commodity
  thermoplastics through dioxaborolane metathesis}.
\newblock \bibinfo{journal}{Science} \bibinfo{volume}{356},
  \bibinfo{pages}{62--65}.
\bibitem[{Schlichting et~al.(2010)Schlichting, de~Andrada, Vieira,
  de~Oliveira~Barra and Magne}]{schlichting2010composite}
\bibinfo{author}{Schlichting, L.H.}, \bibinfo{author}{de~Andrada, M.A.C.},
  \bibinfo{author}{Vieira, L.C.C.}, \bibinfo{author}{de~Oliveira~Barra, G.M.},
  \bibinfo{author}{Magne, P.}, \bibinfo{year}{2010}.
\newblock \bibinfo{title}{Composite resin reinforced with pre-tensioned glass
  fibers. influence of prestressing on flexural properties}.
\newblock \bibinfo{journal}{Dental materials} \bibinfo{volume}{26},
  \bibinfo{pages}{118--125}.
\bibitem[{Shah et~al.(2020)Shah, Patil, Deshpande, Krieg, Kashmari, Al~Mahmud,
  King, Odegard and Maiaru}]{shah2020multiscale}
\bibinfo{author}{Shah, S.}, \bibinfo{author}{Patil, S.},
  \bibinfo{author}{Deshpande, P.}, \bibinfo{author}{Krieg, A.},
  \bibinfo{author}{Kashmari, K.}, \bibinfo{author}{Al~Mahmud, H.},
  \bibinfo{author}{King, J.}, \bibinfo{author}{Odegard, G.M.},
  \bibinfo{author}{Maiaru, M.}, \bibinfo{year}{2020}.
\newblock \bibinfo{title}{Multiscale modeling for virtual manufacturing of
  thermoset composites}, in: \bibinfo{booktitle}{AIAA Scitech 2020 Forum}, p.
  \bibinfo{pages}{0882}.
\bibitem[{Shah et~al.(2023)Shah, Patil, Hansen, Odegard and
  Maiar{\`u}}]{shah2023process}
\bibinfo{author}{Shah, S.P.}, \bibinfo{author}{Patil, S.U.},
  \bibinfo{author}{Hansen, C.J.}, \bibinfo{author}{Odegard, G.M.},
  \bibinfo{author}{Maiar{\`u}, M.}, \bibinfo{year}{2023}.
\newblock \bibinfo{title}{Process modeling and characterization of thermoset
  composites for residual stress prediction}.
\newblock \bibinfo{journal}{Mechanics of Advanced Materials and Structures}
  \bibinfo{volume}{30}, \bibinfo{pages}{486--497}.
\bibitem[{Sharma et~al.(2023)Sharma, Kumar, Rana, Sahoo, Jamil, Kumar, Sharma,
  Li, Kumar, Eldin et~al.}]{sharma2023critical}
\bibinfo{author}{Sharma, H.}, \bibinfo{author}{Kumar, A.},
  \bibinfo{author}{Rana, S.}, \bibinfo{author}{Sahoo, N.G.},
  \bibinfo{author}{Jamil, M.}, \bibinfo{author}{Kumar, R.},
  \bibinfo{author}{Sharma, S.}, \bibinfo{author}{Li, C.},
  \bibinfo{author}{Kumar, A.}, \bibinfo{author}{Eldin, S.M.}, et~al.,
  \bibinfo{year}{2023}.
\newblock \bibinfo{title}{Critical review on advancements on the
  fiber-reinforced composites: Role of fiber/matrix modification on the
  performance of the fibrous composites}.
\newblock \bibinfo{journal}{Journal of Materials Research and Technology} .
\bibitem[{Shi et~al.(2017)Shi, Yu, Kuang, Mu, Dunn, Dunn, Wang and
  Qi}]{shi2017recyclable}
\bibinfo{author}{Shi, Q.}, \bibinfo{author}{Yu, K.}, \bibinfo{author}{Kuang,
  X.}, \bibinfo{author}{Mu, X.}, \bibinfo{author}{Dunn, C.K.},
  \bibinfo{author}{Dunn, M.L.}, \bibinfo{author}{Wang, T.},
  \bibinfo{author}{Qi, H.J.}, \bibinfo{year}{2017}.
\newblock \bibinfo{title}{Recyclable 3d printing of vitrimer epoxy}.
\newblock \bibinfo{journal}{Materials Horizons} \bibinfo{volume}{4},
  \bibinfo{pages}{598--607}.
\bibitem[{Shi et~al.(2021)Shi, Ge, Lu and Yu}]{shi2021nonequilibrium}
\bibinfo{author}{Shi, X.}, \bibinfo{author}{Ge, Q.}, \bibinfo{author}{Lu, H.},
  \bibinfo{author}{Yu, K.}, \bibinfo{year}{2021}.
\newblock \bibinfo{title}{The nonequilibrium behaviors of covalent adaptable
  network polymers during the topology transition}.
\newblock \bibinfo{journal}{Soft Matter} \bibinfo{volume}{17},
  \bibinfo{pages}{2104--2119}.
\bibitem[{Shokrieh et~al.(2014)Shokrieh, Daneshvar and
  Akbari}]{shokrieh2014reduction}
\bibinfo{author}{Shokrieh, M.}, \bibinfo{author}{Daneshvar, A.},
  \bibinfo{author}{Akbari, S.}, \bibinfo{year}{2014}.
\newblock \bibinfo{title}{Reduction of thermal residual stresses of laminated
  polymer composites by addition of carbon nanotubes}.
\newblock \bibinfo{journal}{Materials \& Design} \bibinfo{volume}{53},
  \bibinfo{pages}{209--216}.
\bibitem[{Sorrentino et~al.(2017)Sorrentino, Esposito and
  Bellini}]{sorrentino2017new}
\bibinfo{author}{Sorrentino, L.}, \bibinfo{author}{Esposito, L.},
  \bibinfo{author}{Bellini, C.}, \bibinfo{year}{2017}.
\newblock \bibinfo{title}{A new methodology to evaluate the influence of curing
  overheating on the mechanical properties of thick frp laminates}.
\newblock \bibinfo{journal}{Composites Part B: Engineering}
  \bibinfo{volume}{109}, \bibinfo{pages}{187--196}.
\bibitem[{Tabatabaeian et~al.(2022)Tabatabaeian, Ghasemi, Shokrieh, Marzbanrad,
  Baraheni and Fotouhi}]{tabatabaeian2022residual}
\bibinfo{author}{Tabatabaeian, A.}, \bibinfo{author}{Ghasemi, A.R.},
  \bibinfo{author}{Shokrieh, M.M.}, \bibinfo{author}{Marzbanrad, B.},
  \bibinfo{author}{Baraheni, M.}, \bibinfo{author}{Fotouhi, M.},
  \bibinfo{year}{2022}.
\newblock \bibinfo{title}{Residual stress in engineering materials: a review}.
\newblock \bibinfo{journal}{Advanced engineering materials}
  \bibinfo{volume}{24}, \bibinfo{pages}{2100786}.
\bibitem[{Taynton et~al.(2016)Taynton, Ni, Zhu, Yu, Loob, Jin, Qi and
  Zhang}]{taynton2016repairable}
\bibinfo{author}{Taynton, P.}, \bibinfo{author}{Ni, H.}, \bibinfo{author}{Zhu,
  C.}, \bibinfo{author}{Yu, K.}, \bibinfo{author}{Loob, S.},
  \bibinfo{author}{Jin, Y.}, \bibinfo{author}{Qi, H.J.},
  \bibinfo{author}{Zhang, W.}, \bibinfo{year}{2016}.
\newblock \bibinfo{title}{Repairable woven carbon fiber composites with full
  recyclability enabled by malleable polyimine networks}.
\newblock \bibinfo{journal}{Advanced Materials} \bibinfo{volume}{28},
  \bibinfo{pages}{2904--2909}.
\bibitem[{Winne et~al.(2019)Winne, Leibler and Du~Prez}]{winne2019dynamic}
\bibinfo{author}{Winne, J.M.}, \bibinfo{author}{Leibler, L.},
  \bibinfo{author}{Du~Prez, F.E.}, \bibinfo{year}{2019}.
\newblock \bibinfo{title}{Dynamic covalent chemistry in polymer networks: a
  mechanistic perspective}.
\newblock \bibinfo{journal}{Polymer Chemistry} \bibinfo{volume}{10},
  \bibinfo{pages}{6091--6108}.
\bibitem[{Xu et~al.(2013)Xu, Zhao and Qiao}]{xu2013critical}
\bibinfo{author}{Xu, J.}, \bibinfo{author}{Zhao, Q.}, \bibinfo{author}{Qiao,
  P.}, \bibinfo{year}{2013}.
\newblock \bibinfo{title}{A critical review on buckling and post-buckling
  analysis of composite structures}.
\newblock \bibinfo{journal}{Frontiers in Aerospace Engineering}
  \bibinfo{volume}{2}, \bibinfo{pages}{157--168}.
\bibitem[{Yang et~al.(2018)Yang, Terentjev, Wei and Ji}]{yang2018solvent}
\bibinfo{author}{Yang, Y.}, \bibinfo{author}{Terentjev, E.M.},
  \bibinfo{author}{Wei, Y.}, \bibinfo{author}{Ji, Y.}, \bibinfo{year}{2018}.
\newblock \bibinfo{title}{Solvent-assisted programming of flat polymer sheets
  into reconfigurable and self-healing 3d structures}.
\newblock \bibinfo{journal}{Nature communications} \bibinfo{volume}{9},
  \bibinfo{pages}{1906}.
\bibitem[{Yu et~al.(2016a)Yu, Shi, Dunn, Wang and Qi}]{yu2016carbon}
\bibinfo{author}{Yu, K.}, \bibinfo{author}{Shi, Q.}, \bibinfo{author}{Dunn,
  M.L.}, \bibinfo{author}{Wang, T.}, \bibinfo{author}{Qi, H.J.},
  \bibinfo{year}{2016}a.
\newblock \bibinfo{title}{Carbon fiber reinforced thermoset composite with near
  100\% recyclability}.
\newblock \bibinfo{journal}{Advanced functional materials}
  \bibinfo{volume}{26}, \bibinfo{pages}{6098--6106}.
\bibitem[{Yu et~al.(2016b)Yu, Shi, Li, Jabour, Yang, Dunn, Wang and
  Qi}]{yu2016interfacial}
\bibinfo{author}{Yu, K.}, \bibinfo{author}{Shi, Q.}, \bibinfo{author}{Li, H.},
  \bibinfo{author}{Jabour, J.}, \bibinfo{author}{Yang, H.},
  \bibinfo{author}{Dunn, M.L.}, \bibinfo{author}{Wang, T.},
  \bibinfo{author}{Qi, H.J.}, \bibinfo{year}{2016}b.
\newblock \bibinfo{title}{Interfacial welding of dynamic covalent network
  polymers}.
\newblock \bibinfo{journal}{Journal of the Mechanics and Physics of Solids}
  \bibinfo{volume}{94}, \bibinfo{pages}{1--17}.
\bibitem[{Zhang et~al.(2014a)Zhang, Gu, Wang, Li and Zhang}]{zhang2014effects}
\bibinfo{author}{Zhang, K.}, \bibinfo{author}{Gu, Y.}, \bibinfo{author}{Wang,
  S.}, \bibinfo{author}{Li, M.}, \bibinfo{author}{Zhang, Z.},
  \bibinfo{year}{2014}a.
\newblock \bibinfo{title}{Effects of preheating temperature of the mould on the
  properties of rapid-curing carbon fibre composites fabricated by
  vacuum-assisted resin infusion moulding}.
\newblock \bibinfo{journal}{Polymers and Polymer Composites}
  \bibinfo{volume}{22}, \bibinfo{pages}{825--836}.
\bibitem[{Zhang et~al.(2014b)Zhang, Gu, Zhang et~al.}]{zhang2014effect}
\bibinfo{author}{Zhang, K.}, \bibinfo{author}{Gu, Y.}, \bibinfo{author}{Zhang,
  Z.}, et~al., \bibinfo{year}{2014}b.
\newblock \bibinfo{title}{Effect of rapid curing process on the properties of
  carbon fiber/epoxy composite fabricated using vacuum assisted resin infusion
  molding}.
\newblock \bibinfo{journal}{Materials \& Design (1980-2015)}
  \bibinfo{volume}{54}, \bibinfo{pages}{624--631}.
\bibitem[{Zhao et~al.(2006)Zhao, Hayes, Patterson and Jones}]{zhao2006phase}
\bibinfo{author}{Zhao, F.}, \bibinfo{author}{Hayes, S.},
  \bibinfo{author}{Patterson, E.}, \bibinfo{author}{Jones, F.},
  \bibinfo{year}{2006}.
\newblock \bibinfo{title}{Phase-stepping photoelasticity for the measurement of
  interfacial shear stress in single fibre composites}.
\newblock \bibinfo{journal}{Composites Part A: Applied Science and
  Manufacturing} \bibinfo{volume}{37}, \bibinfo{pages}{216--221}.
\bibitem[{Zhao et~al.(2019)Zhao, Zhou, Wang, Wu, Wang and Chen}]{zhao2019self}
\bibinfo{author}{Zhao, G.}, \bibinfo{author}{Zhou, Y.}, \bibinfo{author}{Wang,
  J.}, \bibinfo{author}{Wu, Z.}, \bibinfo{author}{Wang, H.},
  \bibinfo{author}{Chen, H.}, \bibinfo{year}{2019}.
\newblock \bibinfo{title}{Self-healing of polarizing films via the synergy
  between gold nanorods and vitrimer}.
\newblock \bibinfo{journal}{Advanced Materials} \bibinfo{volume}{31},
  \bibinfo{pages}{1900363}.
\bibitem[{Zheng et~al.(2022)Zheng, Zhang, Li, Zhu, Wang, Song, Wu, Yang, Huang
  and Ma}]{zheng2022recent}
\bibinfo{author}{Zheng, H.}, \bibinfo{author}{Zhang, W.}, \bibinfo{author}{Li,
  B.}, \bibinfo{author}{Zhu, J.}, \bibinfo{author}{Wang, C.},
  \bibinfo{author}{Song, G.}, \bibinfo{author}{Wu, G.}, \bibinfo{author}{Yang,
  X.}, \bibinfo{author}{Huang, Y.}, \bibinfo{author}{Ma, L.},
  \bibinfo{year}{2022}.
\newblock \bibinfo{title}{Recent advances of interphases in carbon
  fiber-reinforced polymer composites: A review}.
\newblock \bibinfo{journal}{Composites Part B: Engineering}
  \bibinfo{volume}{233}, \bibinfo{pages}{109639}.

\end{thebibliography}

\end{document}